\documentclass[prapplied,twocolumn,showpacs,preprintnumbers,amsmath,amssymb]{revtex4}
\usepackage{graphicx}
\usepackage{longtable}
\usepackage{dcolumn}
\usepackage{bm}
\usepackage{color}

\usepackage[colorlinks=true, urlcolor=blue,  linkcolor=blue,  citecolor=blue]{hyperref}
\usepackage[all]{hypcap}

\begin{document}
	


\title{Proposal for semiconductor-free negative differential resistance tunnel diode \\  with ultra-high peak-to-valley current ratio}

\author{Ersoy \c{S}a\c{s}{\i}o\u{g}lu$^{1}$}\email{ersoy.sasioglu@physik.uni-halle.de}
\author{Ingrid Mertig$^{1,2}$}

\affiliation{$^{1}$Institute of Physics, Martin Luther University Halle-Wittenberg, 06120 Halle (Saale), Germany \\
$^{2}$Max Planck Institute of Microstructure Physics, Weinberg 2, 06120 Halle (Saale), Germany}

\date{\today}

\begin{abstract}

The negative differential resistance (NDR) tunnel diodes are promising alternative devices 
for beyond-CMOS computing as they offer several potential applications when integrated with 
transistors. We propose a novel semiconductor-free  NDR tunnel diode concept that exhibits 
ultra-high peak-to-valley current ratio (PVCR) value. Our proposed NDR diode consists of two 
cold metal  electrodes separated by a thin insulating tunnel barrier. The NDR effect stems 
from the unique electronic band structure of the cold metal electrodes, i.e., the width of 
the isolated metallic bands around the Fermi level as well as the energy gaps separating 
higher- and lower-lying bands determine the current-voltage ($I$-$V$) characteristics 
and the PVCR value of the tunnel diode. By proper choice of the  cold metal electrode 
materials, either a conventional N-type or $\Lambda$-type NDR effect can be obtained. 
Two-dimensional (2D) nanomaterials offer a unique platform for the realization of proposed 
NDR tunnel diodes. To demonstrate the proof of concept we employ the nonequilibrium Green 
function method combined with density functional theory to calculate the $I$-$V$ characteristic 
of the lateral (AlI$_2$/MgI$_2$/AlI$_2$) and vertical (NbS$_2$/h-BN/NbS$_2$)  heterojunction
tunnel diodes based on 2D cold metals. For the lateral tunnel diode, we obtain a  $\Lambda$-type 
NDR effect with an ultra-high PVCR value of 10$^{16}$ at room temperature, while the vertical 
tunnel diode exhibits a conventional N-type NDR effect with a smaller PVCR value of about 
10$^4$. The proposed concept provides a semiconductor-free solution for NDR devices to 
achieve desired $I$-$V$ characteristics with ultra-high PVCR values for memory and logic 
applications.

\end{abstract}



\maketitle

\section{Introduction}

Current logic and memory devices are based on the complementary metal oxide semiconductor
(CMOS) field-effect transistor technology and more than forty years chip makers have 
succeeded in scaling of CMOS transistors, which allowed Moore’s law to remain on track\cite{Moore}. 
However, the 2D scaling of the CMOS technology will reach fundamental limits soon according 
to the International Roadmap for Devices and Systems\cite{IRDS}. In future technology nodes the 
performance benefits of CMOS-based devices might no longer meet the circuit performance targets 
and thus require a hybrid circuit solution. Within the last two decades several beyond-CMOS 
device concepts ranging from electronic to spintronic, from ferroelectric to resistive switching 
devices have been proposed and demonstrated 
\cite{chen2014emerging,ionescu2011tunnel,nikonov2011proposal,amlani1999digital,allwood2005magnetic}. 
Despite many valuable features of these new devices, extensive benchmark calculations 
have shown that none of them can beat CMOS transistors \cite{pan2017expanded}. Thus, most of 
the beyond-CMOS devices aim to complement CMOS rather than replace it and they might enable 
computing paradigms beyond the capabilities of conventional CMOS technology.

The negative differential resistance (NDR) tunnel diodes offer variety of unique functionalities
and potential applications when integrated with conventional CMOS transistors\cite{berger2011negative}.
Among them, NDR-based multi-valued logic gates and static random-access-memory (SRAM) are the most 
promising applications \cite{jo2021recent,van1999tunneling}. The NDR effect allows to 
design logic and memory architectures with much reduced device count, very high speed, and much lower 
power consumption. For instance, a tunnel SRAM requires a single transistor and two NDR tunnel 
diodes instead of six transistors in conventional SRAM architecture, which substantially reduces the 
device footprint area and power consumption \cite{van1999tunneling,karda2009one}. By employing two 
NDR diodes connected in series the monostable-bistable transition logic element gate has
been demonstrated to perform  NAND and NOR operations \cite{chen1996inp,maezawa1998high,williamson199712}. 
All other logic gates have also been implemented by combining NDR diodes with CMOS transistors
\cite{williamson199712}. The binary NDR-based logic has also been extended to the multi-valued logic, 
which  increases the information density per given unit device and decreases the overall system complexity 
\cite{micheel1990differential,lin1994resonant,jin2004tri}. By shifting from binary to the 
ternary logic the system complexity can be drastically reduced to 63\% \cite{hurst1984multiple}.

\begin{figure}[!t]
\includegraphics[scale=0.15]{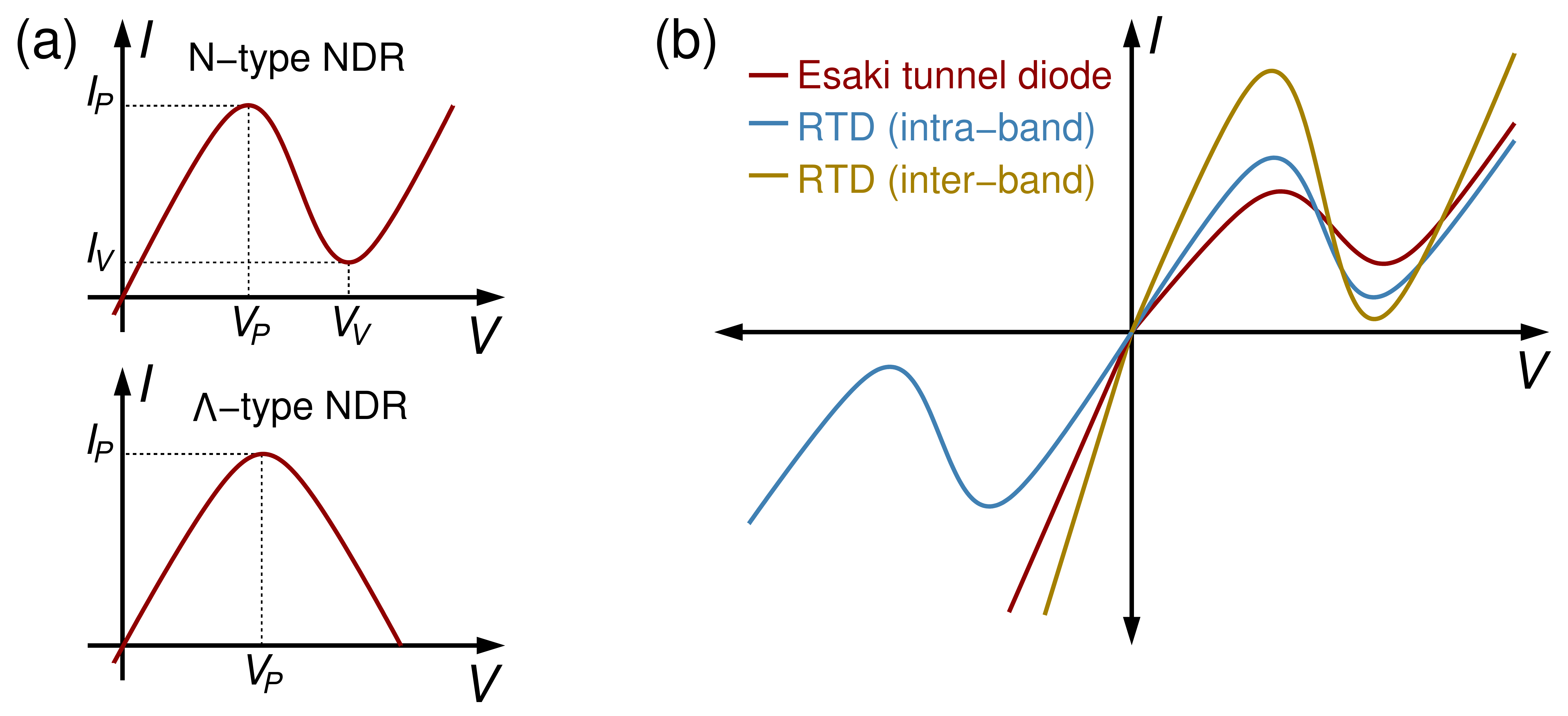}
\vspace{-0.55 cm}
\caption{(Color online) (a) N-type and $\Lambda$-type  negatives differential resistance (NDR). 
(b) Schematic representation of the current-voltage ($I$-$V$) characteristics of a Esaki tunnel diode 
and intra- and inter-band resonant-tunneling diodes.  }
\label{fig1}
\end{figure}

There are several devices (two-terminal tunnel diodes) and circuits described in the literature 
that produce an NDR effect. Of the two-terminal tunnel diodes, the Esaki diode and the resonant 
tunneling diode received the most attention \cite{esaki1958new,esaki1970superlattice,ramesh2012high}. 
The Esaki diode is a heavily doped (degenerate) p-n junction diode, in which the electron transport 
in the contact region is via quantum mechanical tunneling under forward bias and it shows NDR effect, 
i.e, electrical current decreases with increasing bias voltage. Figure\,\ref{fig1} schematically 
illustrates the current-voltage ($I$-$V$) characteristics of an Esaki diode and an intra- and 
inter-band resonant tunneling diodes. However, those two-terminal NDR diodes have two primary 
features hindering their applications: i) the peak-to-valley current ratio (PVCR) is rather low 
in tunnel diodes that are compatible with CMOS technology 
\cite{ramesh2012high,fung2011esaki,duschl1999high,schmid2012silicon}. In these diodes the PVCR is
usually less than ten, not high enough for memory applications (tunnel SRAM). ii) the III-V 
semiconductors that produce high PVCR (between 5 and 144) are not compatible with current CMOS 
technology \cite{chow1992investigation,tsai1994pn}. The low PVCR values in two-terminal NDR tunnel 
diodes have been attributed to the band tails tunneling, which originates from the strong doping 
and doping fluctuations and it has been studied in detail by many authors using different approaches 
\cite{sant2017effect,bizindavyi2018band,schenk2020tunneling}. Many research contributions have 
been reported to improve the PVCR value  over 100 based on a CMOS compatible process. Extremely 
high, by 2-3 orders of magnitude, PVCR values have been obtained in NDR circuits based on combination 
of a Si-based CMOS and a SiGe heterojunction-bipolar-transistor \cite{chung2004three,chen2009negative,gan2007fabrication,duane2003bistable,fang2022giant}. 
Despite their high PVCR values, such devices posses complex circuit topology with at least three (four) 
transistors for a $\Lambda$-type (N-type) NDR effect (see Figure\,\ref{fig1}), 
which makes them unsuitable for memory applications \cite{gan2016multiple}.

\begin{figure}[!t]
\includegraphics[scale=0.15]{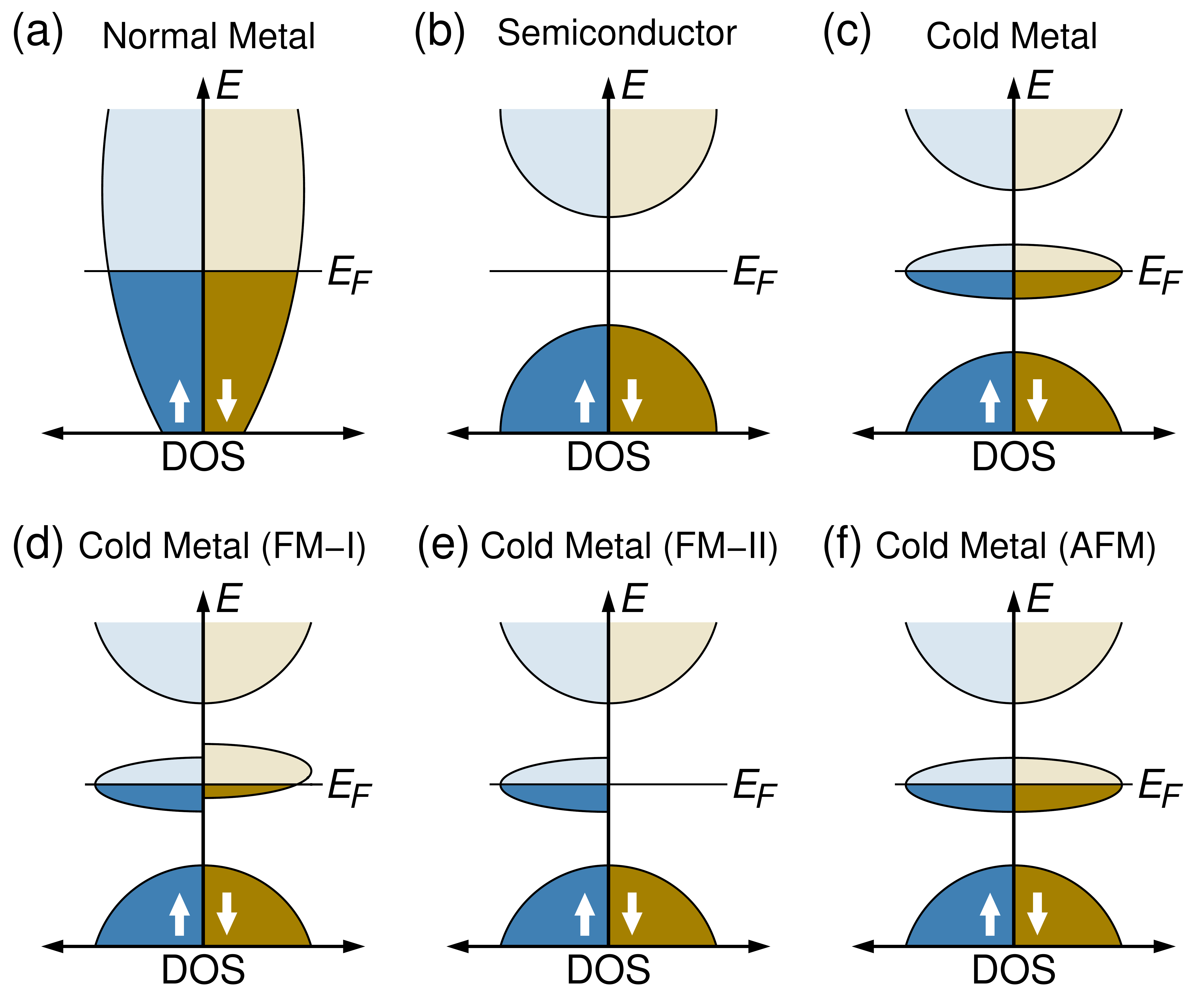}
\vspace{-0.2 cm}
\caption{(Color online) Schematic representation of the density of states (DOS) for
(a) a normal metal, (b) a  semiconductor, (c) a non-magnetic (or paramagnetic) cold-metal, 
(d) a ferromagnetic (FM-I) cold-metal, (e) a half-metallic ferromagnetic (FM-II) cold-metal, 
and (f) an antiferromagnetic (AFM) cold-metal.}
\label{fig2}
\end{figure}

Recently, we proposed a new semiconductor-free NDR tunnel diode concept with ultra-high 
PVCR. A patent application has been ﬁled for the proposed device \cite{sasioglu2021patent} 
and preliminary results have been presented in Ref.\cite{sasioglu2022negative} Our proposed 
NDR diode consists of two cold metal electrodes separated by a thin insulating tunnel 
barrier. In Figure\,\ref{fig2} we schematically show the density of states (DOS) of a cold 
metal and compare with a normal metal and semiconductor.  The NDR effect stems from the 
unique electronic band structure of the cold metal electrodes, i.e., the width of the 
isolated metallic bands around the Fermi level as well as the energy gaps separating 
higher and lower lying bands determine the current-voltage ($I$-$V$) characteristics 
and the PVCR value of the tunnel diode.  By proper choice of the cold metal electrode 
materials either conventional N-type or $\Lambda$-type NDR effect can be obtained. In 
this paper we provide a detailed description of the new NDR tunnel diode concept and 
demonstrate the proof of principle by \textit{ab initio} quantum transport calculations. 
Furthermore, we address diﬀerent classes of suitable materials for realizing the new 
concept. To demonstrate both $\Lambda$-type and N-type NDR effects we choose two different 
kinds of cold metal materials: AlI$_2$ and  NbS$_2$ (see Table\,\ref{table}). The former 
posses a narrow metallic band around the Fermi level and relatively large energy gaps 
below and above the Fermi energy, while for the latter material the width of the metallic 
band and energy gaps are more or less comparable. By employing the nonequilibrium Green 
function (NEGF) method combined with density functional theory (DFT) we have calculated 
the $I$-$V$ curves of the lateral AlI$_2$/MgI$_2$/AlI$_2$ and vertical NbS$_2$/h-BN/NbS$_2$ 
heterojunction tunnel diodes. We obtain a $\Lambda$-type NDR effect for the lateral tunnel 
diode with an ultra-high  PVCR value of 10$^{16}$ at room temperature, while the vertical 
tunnel diode exhibits conventional N-type NDR effect with a smaller PVCR value of 10$^{4}$
within the coherent tunneling transport model.  

\section{Device concept}
\label{section2}

\begin{figure*}[!t]
\includegraphics[scale=0.245]{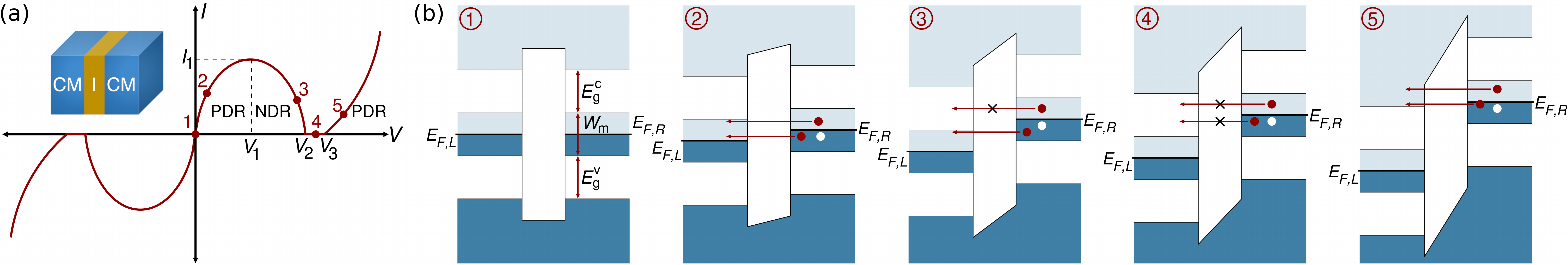}
\vspace{-0.4cm}
\caption{(Color online) (a) Schematic representation of the cold metal NDR tunnel diode  and 
the corresponding current-voltage ($I$-$V$) characteristics. (b) A schematic representation of 
the band diagram and thermal electron-hole excitations for the cold metal NDR tunnel diode for 
different forward bias voltages.}
\label{fig3}
\end{figure*}

The structure of the proposed  cold metal tunnel diode and its $I$-$V$ characteristics are 
shown schematically in Figure\,\ref{fig3}a. Analogous to the metal-insulator-metal diode, 
the NDR diode consists of two cold metal electrodes separated by a thin insulating tunnel 
barrier. Cold metals possess unique electronic band structure (see Figure\,\ref{fig3} and 
Figure\,\ref{fig4}), i.e., isolated metallic bands around the Fermi energy are separated 
from the higher-lying and lower-lying states by sizeable energy gaps. The $I$-$V$ characteristics
of the cold metal tunnel diode can be qualitatively explained on the basis of the schematic 
energy-band diagram shown in Figure\,\ref{fig3}b, where the energy-band diagram of the insulating 
tunnel barrier is represented by a rectangular potential without bias voltage and by a trapezoid 
when a ﬁnite bias is applied. The metallic band width and energy gaps below and above the 
Fermi level are denoted by $W_{\mathrm{m}}$, $E_{\mathrm{g}}^{\mathrm{V}}$, and $E_{\mathrm{g}}^{\mathrm{C}}$, 
respectively. In energy-band diagram (Figure\,\ref{fig3}b) we assume that $W_{\mathrm{m}}=E_{\mathrm{g}}^{\mathrm{V}}=E_{\mathrm{g}}^{\mathrm{C}}$.
If we choose the same cold metal materials for the left and right electrodes of the tunnel 
diode, then no charge transfer takes place between the electrodes.

For a qualitative discussion of the NDR diode characteristics we will use Bardeen’s approach for
tunneling \cite{bardeen1961tunnelling,meservey1994spin}. For a simple tunnel barrier the tunneling 
current $I(V)$ for a bias voltage $V$ is given by the expression $I(V)  \sim  \int_{-\infty}^{+\infty} 
\rho_{\textrm{CM}}(E+eV) \rho_{\textrm{CM}}(E) |T(V)|^2  f(E)[1-f(E+eV)]dE$, where 
$\rho_{\textrm{CM}}(E+eV)$  denotes  the density of states of the  cold metal electrode and $f(E)$ 
being the Fermi distribution function. $T(V)$ is the transmission probability being proportional 
to $e^{-d\sqrt{\phi-V}}$, where $d$ is the thickness of the tunnel barrier and $\phi$ 
being the barrier height. As shown in Figure\,\ref{fig3}b when a positive bias voltage is 
applied to the left cold metal electrode (second panel in Figure\,\ref{fig3}b), electrons from 
the occupied valence band of the right cold metal  electrode tunnel through the insulating 
barrier into the unoccupied states of the left electrode. With increasing bias voltage $V$ the
current $I$ increases up to a peak value $I_1$ then it starts to decrease (panel 3) and finally 
for a specific value of bias voltage $V_2$ the current becomes zero. This stems from 
the fact that the conduction band of the left cold metal electrode has an energy gap 
$E_{\mathrm{g}}^{\mathrm{C}}$ and thus within the applied potential window there are no 
available states for the electrons to tunnel into (see panel 4), resulting in a zero current. At 
zero temperature this current is exactly zero, while at finite temperature very few thermally 
excited high energy electrons that are at the tail of the Fermi-Dirac distribution of the right 
cold metal electrode can tunnel into the left electrode giving rise to an extremely small current 
and thus the PVCR takes a finite value instead of a infinite value at room temperature. The current 
remains zero between the bias potential $V_2$ and $V_3$ and after $V_3$ the current starts to 
increase again. The region between the bias potential $V_1$ and $V_2$ is called the NDR region. 
The current in the first positive differential resistance (PDR) region and in NDR region (see 
Figure\,\ref{fig3}a) stems from the intra-band tunneling, while in the second PDR region it is 
due to the inter-band tunneling (panel 5 in Figure\,\ref{fig3}b). Note that the applied bias voltage 
drops across the insulating tunnel barrier.

The $I$-$V$ characteristics of a  cold metal tunnel diode having the same left and right electrodes 
is mainly determined by three parameters  $W_{\mathrm{m}}$, $E_{\mathrm{g}}^{\mathrm{V}}$, 
and $E_{\mathrm{g}}^{\mathrm{C}}$. For N-type NDR effect one of the the following three conditions 
should be satisfied: i) $W_{\mathrm{m}} \sim E_{\mathrm{g}}^{\mathrm{V}} \sim E_{\mathrm{g}}^{\mathrm{C}}$, 
ii) $W_{\mathrm{m}} \sim E_{\mathrm{g}}^{\mathrm{C}} < E_{\mathrm{g}}^{\mathrm{V}}$, 
iii) $W_{\mathrm{m}} \sim E_{\mathrm{g}}^{\mathrm{V}} < E_{\mathrm{g}}^{\mathrm{C}}$. In this case 
$ V_2 \simeq V_3$ in $I$-$V$ characteristics shown in Figure\,\ref{fig3}a. When none of these 
three conditions are satisfied the NDR effect can still be observed but with a substantially 
reduced PVCR value. For a $\Lambda$-type NDR effect the metallic band width should be much smaller 
than the both  energy gaps, i.e., $W_{\mathrm{m}} \ll E_{\mathrm{g}}^{\mathrm{V}} \sim E_{\mathrm{g}}^{\mathrm{C}}$  
and this leads to $V_2 \ll V_3$ in $I$-$V$ characteristics. For both types of NDR effects the bias 
voltages $V_1$ and $V_2$ corresponding  to the peak and valley currents are approximately determined by 
$W_{\mathrm{m}}/2$ and $W_{\mathrm{m}}$, respectively.

Up to now we focus only on the forward bias and discuss the NDR effect. In the same way, for 
a reverse bias one can observe the same NDR effect, i.e., the $I$-$V$ characteristics of the 
cold metal tunnel diode turns out to be  anti-symmetric if the left and right electrodes are 
made of the same materials. This is similar to the intra-band resonant tunneling diode but 
different than the Esaki diode and inter-band resonant tunneling diode shown in Figure\,\ref{fig1}b. 
As the $I$-$V$ characteristics of the cold metal tunnel diode mainly depends on the three 
electronic structure parameters $W_{\mathrm{m}}$, $E_{\mathrm{g}}^{\mathrm{V}}$, and 
$E_{\mathrm{g}}^{\mathrm{C}}$, by proper choice of the left and right electrode materials one 
can achieve desired $I$-$V$ characteristics. For instance, the NDR region ($V_1$-$V_2$ interval) 
can be tuned by tuning these parameters. Moreover, multi-peak NDR effect can be achieved in 
cold metal tunnel diodes connected in series.

Cold metallic behaviour is rather rare among the three-dimensional (3D) bulk materials. Known 
examples are alkali-doped fullerides \cite{gunnarsson2004alkali}, high temperature paramagnetic 
phase of V$_2$O$_3$ \cite{hansmann2013mott}, and recently several flat-band materials such 
as CoAl$_2$O$_4$, Rb$_2$NbCl$_6$, Sr$_3$PbNiO$_6$, etc  have been predicted to show cold 
metallic behaviour \cite{regnault2022catalogue}. On the other hand, two-dimensional (2D) 
transition-metal dichalcogenides (NbX$_2$, TaX$_2$, X=S, Se, Te) and transition-metal dihalides 
(ScI$_2$, YI$_2$, etc) as well as several other 2D materials like AlI$_2$, GaI$_2$, InI$_2$, 
Ag$_2$F$_4$, DySBr, DySI, NdOBr, SmOBr, etc exhibit cold-metallic behaviour 
\cite{haastrup2018computational,gjerding2021recent,mounet2018two} and offer a unique platform 
for the realization of the proposed NDR tunnel diodes. Moreover, in contrast to 3D materials, 
2D materials are much better suited for tunneling devices as they form high-quality 
heterointerfaces due to the absence of dangling bonds. Recently conventional N-type NDR effect 
has been observed in a variety of  devices based on 1D\cite{janatipour2022achieved,ostovan2018length}
and 2D materials, such as graphene/h-BN/graphene resonant tunnel diodes,\cite{mishchenko2014twist} 
black phosphorus tunnel diodes \cite{shim2016phosphorene,kim2020multiple}, as well as transition-metal 
dichalcogenide tunnel diodes \cite{lin2015atomically,roy2015dual,duong2018parameter}. However, 
in all these 2D material based NDR tunnel devices the PVCR values were found to be less than 10 
and thus they do not exhibit their theoretical advantages. As will be shown in the following 
section the tunnel diodes based on 2D cold metals overcome the limitations of 
conventional semiconductor tunnel devices by providing ultra-high PVCR values as well as both 
types of NDR effects, i.e., conventional N-type NDR and $\Lambda$-type NDR effect. The latter 
can be achieved only in multi-terminal NDR circuits.

\section{Computational Methods}
\label{section3}

Ground state electronic structure calculations are carried out using DFT implemented in the 
QuantumATK S-2021.06 package\cite{smidstrup2019an}. As basis-set we use linear combinations 
of atomic orbitals (LCAO)  together with norm-conserving PseudoDojo pseudopotentials\cite{QuantumATKPseudoDojo} 
with the Perdew-Burke-Ernzerhof (PBE) parametrization of the exchange-correlation functional\cite{perdew1996generalized}. We use DFT-D2 van der Waals corrections for the 
vertical tunnel diode.  A dense $24 \times 24 \times 1$ $\mathbf{k}$-point grid and density 
mesh cutoff of 120\,Hartree have been used. To prevent interactions between the periodically 
repeated images, 20\,{\AA} of vacuum were added and Dirichlet and Neumann boundary conditions 
are employed to the upper and lower surface, respectively. The total energy and forces have 
been converged at least to 10$^{-4}$\,eV and 0.01\,eV/{\AA}, respectively.

The transport calculations were performed using DFT combined with the NEGF method. We use a
$24 \times 1 \times 172$ ($8 \times 1 \times 172$) $\mathbf{k}$-point grid in self-consistent 
DFT-NEGF calculations of lateral (vertical) tunnel diodes. The $I$-$V$ characteristics were 
calculated within a Landauer approach~\cite{Landauer-Buettiker}, where $ I(V) = \frac{2e}{h}\int 
\,T(E,V)\left[f_{L}(E,V)-f_{R}(E,V)\right] \mathrm{d}E $. Here $V$ denotes the bias 
voltage, $T(E,V)$ is the transmission coefficient and $f_L(E,V)$ and $f_R(E,V)$ are the 
Fermi-Dirac distributions of the left and right electrodes, respectively. The transmission 
coefficient $T(E,V)$ for lateral (vertical) tunnel diode is calculated using a $300 
\times 1$ ($100 \times 1$) $\mathbf{k}$-point grid.

\section{Results and Discussion}
\label{section4}

\begin{table*}[!ht]
\caption{\label{table}
Material composition of the tunnel diode, type of NDR effect, crystal structure, 
lattice constants $a$,  PBE (PBE-1/2) band gap $E_{\mathrm{g}}$ of the insulating tunnel 
barrier, PBE (PBE-1/2) metallic band width $W_{\mathrm{m}}$, and energy gaps 
$E_{\mathrm{g}}^{\mathrm{V}}$ and $E_{\mathrm{g}}^{\mathrm{C}}$ below  and above the 
Fermi level for the cold metallic electrode materials. Lattice parameters are taken from Ref.\cite{haastrup2018computational}.}
\begin{ruledtabular}
\begin{tabular}{lllccccc} 
Tunnel Diode & NDR & Compound & $a$({\AA}) & $E_{\mathrm{g}}$(eV) & $W_{\mathrm{m}}$(eV)  &  $E_{\mathrm{g}}^{\mathrm{V}}$(eV)  & $E_{\mathrm{g}}^{\mathrm{C}}$(eV)    \\ \hline
AlI$_2$/MgI$_2$/AlI$_2$ & $\Lambda$-type & 1T-AlI$_2$  & 4.14           &    & 1.20\,(0.95) & 1.65\,(2.96)  & 1.55\,(2.29)  \\
                        &                & 1T-MgI$_2$  & 4.21 & 3.66\,(5.45) &      &       &       \\
NbS$_2$/BN/NbS$_2$    & N-type         & 1H-NbS$_2$  & 3.34 &              & 1.25\,(1.20) & 0.44\,(1.21)  & 1.36\,(1.38)   \\
                        &                & h-BN        & 2.51 & 4.69\,(6.17) &      &       &         \\
\end{tabular}
\end{ruledtabular}
\end{table*}

\begin{figure*}[!th]
\includegraphics[scale=0.335]{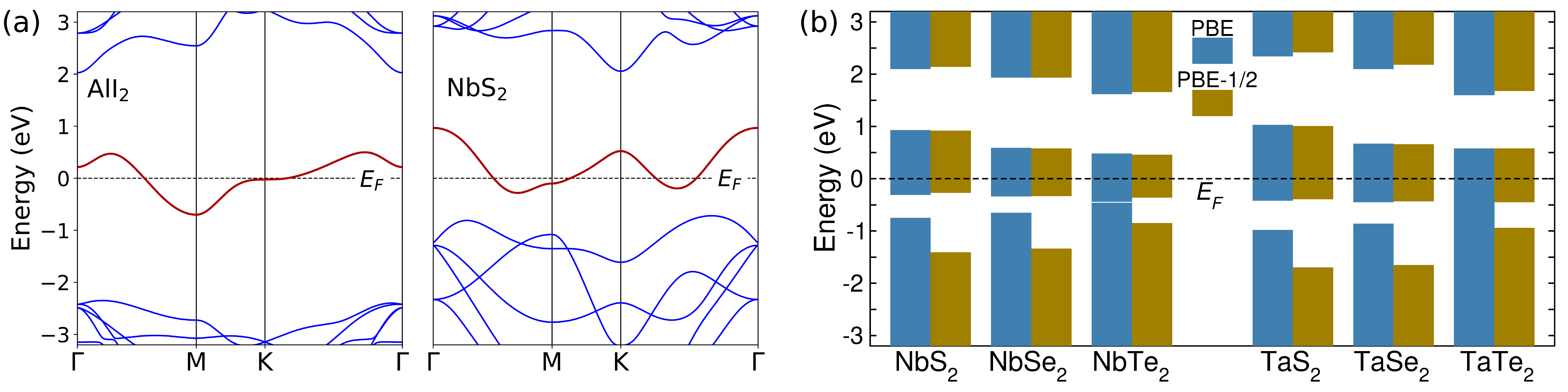}
\vspace{-0.7cm}
\caption{(Color online) (a) Calculated PBE band structure of the monolayer 1T AlI$_2$ and 1H 
NbS$_2$ along the high symmetry lines in 2D Brilloiun zone. The dashed black lines denote the 
Fermi level, which is set to zero. (b) PBE and PBE-1/2 method comparison of the metallic band 
widths ($W_{\mathrm{m}}$), and energy gaps below ($E_{\mathrm{g}}^{\mathrm{V}}$) 
and above ($E_{\mathrm{g}}^{\mathrm{C}}$) the Fermi level for NbX$_2$ and TaX$_2$ 
(X=S, Se, Te) compounds.}
\label{fig4}
\end{figure*}

The cold metal tunnel diode concept introduced in the preceding section can be realized either 
by using 3D materials or 2D van der Waals materials. In the following, due to their structural 
simplicity and high-quality heterointerfaces because of the absence of dangling bonds in vertical 
devices, we will focus on the 2D materials by considering a lateral and a vertical tunnel diode 
as shown in Table\,\ref{table} and demonstrate the proof of principle by \textit{ab-initio} 
quantum transport calculations. As pointed out above, depending on the electronic band structure 
parameters ($W_{\mathrm{m}}$, $E_{\mathrm{g}}^{\mathrm{V}}$, and $E_{\mathrm{g}}^{\mathrm{C}}$) 
of the cold metal electrode materials, a tunnel diode showing either conventional N-type or 
$\Lambda$-type NDR effect can be realized. For this purpose, as electrode materials we choose
AlI$_2$ (1T phase) and NbS$_2$ (1H phase) and DFT-PBE electronic band structure of both materials 
are shown in Figure\,\ref{fig4}a. As seen, both materials have a single metallic band of similar 
width ($W_{\mathrm{m}}$) crossing the Fermi level and an energy gap above the Fermi level
($E_{\mathrm{g}}^{\mathrm{C}}$) of similar size. While the energy  gap $E_{\mathrm{g}}^{\mathrm{V}}$
of AlI$_2$ below the Fermi energy is much larger than the corresponding energy gap of NbS$_2$
(see Table\,\ref{table}). Note that our ground-state calculations for both materials are in good 
agreement with previously published data\cite{haastrup2018computational}. As for the case of 
semiconductors and insulators, the DFT-PBE method underestimates the energy gaps 
($E_{\mathrm{g}}^{\mathrm{V}}$, and $E_{\mathrm{g}}^{\mathrm{C}}$)  of the cold metals under 
consideration. The accurate $E_{\mathrm{g}}^{\mathrm{V}}$ and $E_{\mathrm{g}}^{\mathrm{C}}$ 
values are of particular importance in tunnel diodes made of transition-metal dichalcogenides 
as they possess $E_{\mathrm{g}}^{\mathrm{V}}$ values, which are smaller than the $W_{\mathrm{m}}$ 
in DFT-PBE method, which might result in smaller PVCR values. Taking electronic correlations 
into account either within the HSE06 hybrid functional or $GW$ method results in larger 
$E_{\mathrm{g}}^{\mathrm{V}}$ and  $E_{\mathrm{g}}^{\mathrm{C}}$ values \cite{haastrup2018computational,heil2018quasiparticle,kim2017quasiparticle}. 
For computational purposes, in transport calculations we adopt the PBE-1/2 method, which 
gives rather accurate band gaps in semiconductors and insulator. For instance, PBE gives 
4.69 eV for the band gap of h-BN, an insulator used as a tunnel barrier in vertical 
heterojunction diode, while the PBE-1/2 method more or less reproduces the experimental 
band gap of 6.1 eV (see Table\,\ref{table}). Figure\,\ref{fig4}b shows the $W_{\mathrm{m}}$, 
$E_{\mathrm{g}}^{\mathrm{V}}$, and $E_{\mathrm{g}}^{\mathrm{C}}$ values calculated with the 
PBE and PBE-1/2 methods for one monolayer NbX$_2$ and TaX$_2$ (X=S, Se, Te) compounds. As seen 
in all compounds the $W_{\mathrm{m}}$ values do not change at all in PBE-1/2 method and also the 
change of $E_{\mathrm{g}}^{\mathrm{C}}$ values are negligible. However, PBE-1/2 method gives 
considerably larger $E_{\mathrm{g}}^{\mathrm{V}}$ values, which are  comparable to the ones 
obtained by the HSE06  method \cite{haastrup2018computational} but larger than the $GW$ values 
reported in the literature \cite{heil2018quasiparticle,kim2017quasiparticle}. Note that the $GW$ 
calculations for the electronic band structure of NbS$_2$ is in good agreement with the 
angle-resolved photoemission spectroscopy measurements.\cite{heil2018quasiparticle,PhysRevB.105.245145}

\begin{figure*}[!t]
\includegraphics[scale=0.91]{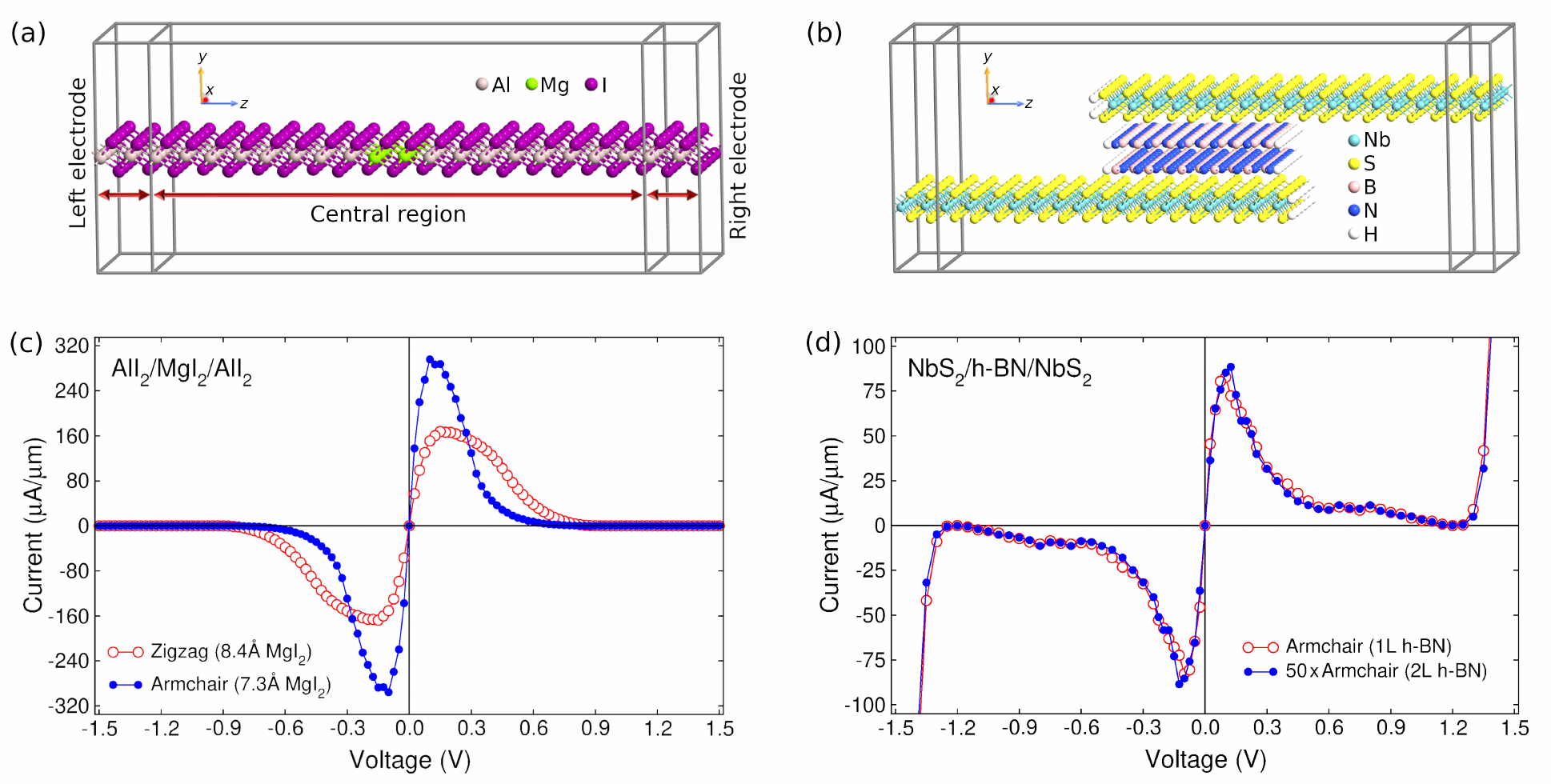}
\vspace{-0.6cm}
\caption{(Color online) (a) The atomic structure of the lateral AlI$_2$/MgI$_2$/AlI$_2$ 
cold metal tunnel diode. The system is periodic along the x-direction orthogonal to the 
z-axis, which is the transport direction. (b) The same as (a) for the vertical NbS$_2$/h-BN/NbS$_2$  
cold metal tunnel diode. (c) The current-voltage characteristics for the lateral 
AlI$_2$/MgI$_2$/AlI$_2$ tunnel diode in both armchair and zigzag orientations. (d) The 
same as (c) for the vertical NbS$_2$/h-BN/NbS$_2$ cold metal tunnel diode in armchair 
orientation.}
\label{fig5}
\end{figure*}

In Figure\,\ref{fig5}a  and \ref{fig5}b, we show the atomic structure of the lateral 
AlI$_2$/MgI$_2$/AlI$_2$ and vertical NbS$_2$/h-BN/NbS$_2$ heterojunction tunnel 
diodes, respectively. The lateral tunnel diode is formed by joining one monolayer of 
AlI$_2$ (left electrode), a one monolayer tunnel barrier MgI$_2$, and one monolayer of 
AlI$_2$ (right electrode). Due to the very similar lattice parameters of electrode 
and tunnel barrier materials (see Table\,\ref{table}) as well as their similar 
compositions, they form a perfect interface. We assume periodicity of the device in 
the $x$-direction. The $z$-direction is chosen as the transport direction. 
We consider both armchair and zigzag orientations and the thickness of tunnel barrier 
is $7.3$\,{\AA} for the former case and $8.4$\,{\AA} for the latter case. The total length 
of the scattering region is about $64$\,{\AA} for both orientations. On the other hand, 
for the vertical tunnel diode the lattice mismatch between NbS$_2$ and tunnel 
barrier h-BN is relatively large and thus in device modelling we construct an orthorhombic 
commensurate supercell of $3\times 3$ for NbS$_2$  and $4\times 4$ for h-BN and consider 
only armchair orientation as sown in Figure\,\ref{fig5}b with an overlap region of about
$24$\,{\AA} and the scattering region is about $70$\,{\AA}.

\begin{figure*}[!t]
\includegraphics[scale=0.265]{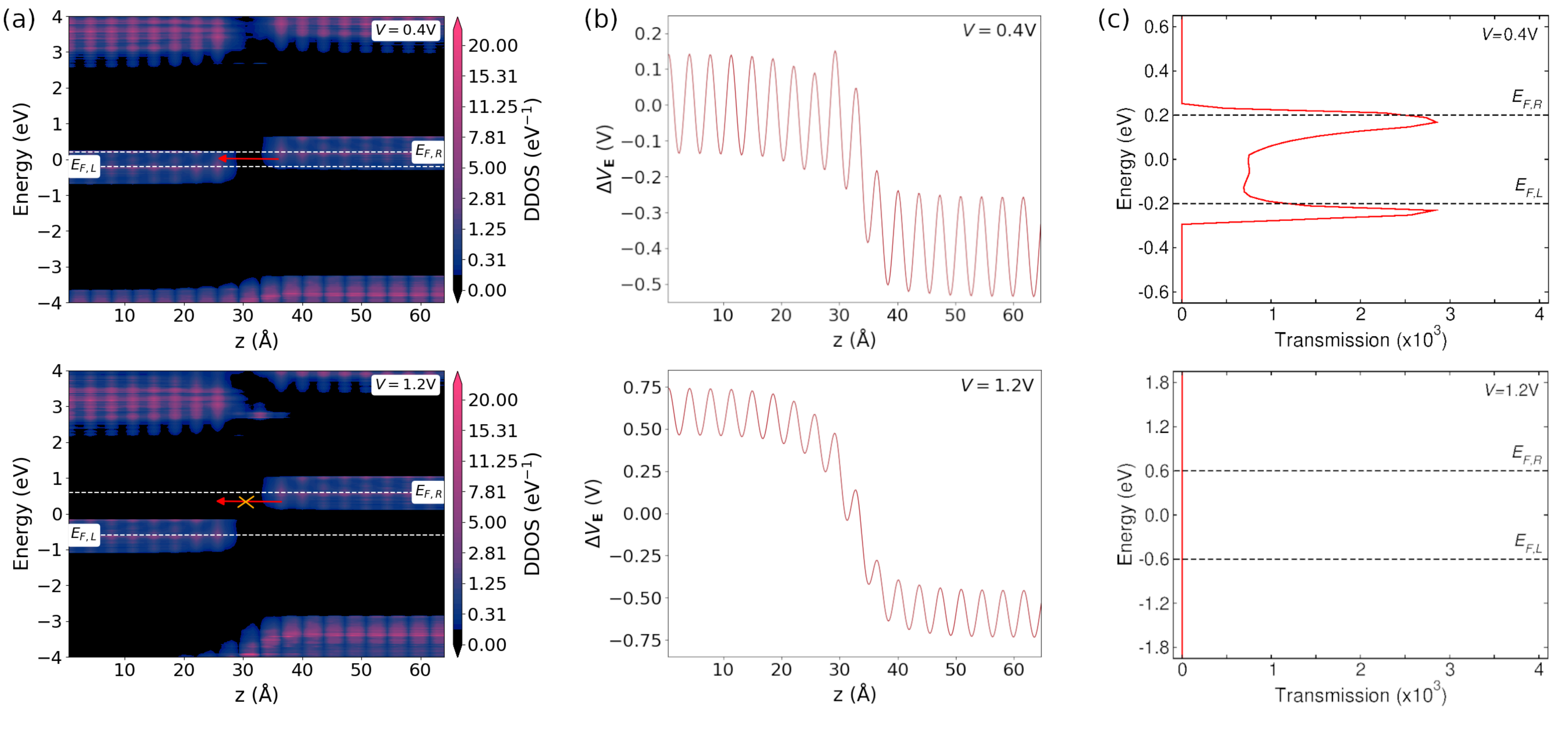}
\vspace{-0.5cm}
\caption{(Color online) (a) Device density of states (DDOS) for the lateral AlI$_2$/MgI$_2$/AlI$_2$ 
cold metal tunnel diode for an applied bias voltage of $V=0.4$\,V (upper panel) and $V=1.2$\,V 
(lower panel) calculated with PBE-1/2 method. The dashed lines display the Fermi level of the left 
and right electrodes (the corresponding atomic structure is presented in Figure\,\ref{fig5}a). 
(b) and (c) calculated electrostatic difference potential and transmission spectrum, respectively.}
\label{fig6}
\end{figure*}

Next, we discuss the transport properties of the cold metal heterojunction tunnel diodes. 
In the preceding section we qualitatively discussed the operation principle and  $I$-$V$ 
characteristics of a  cold metal tunnel diode on the basis of a schematic energy-band 
diagram of the electrodes and a simple tunnel barrier model. However, in real materials 
quantum tunneling is a very sophisticated process as it depends on the symmetry 
of Bloch states in the electrodes, their matching at the interface, and their decay rate, 
which is determined by thickness and barrier height as well as the complex energy bands of 
the tunnel barrier. Thus, fully \textit{ab-initio} atomic scale transport calculations 
are needed to reveal the $I$-$V$ characteristics of the proposed  cold metal tunnel diodes. 
Calculated $I$-$V$ curves are presented in Figure\,\ref{fig5}c and \ref{fig5}d for the lateral 
and the vertical tunnel diode, respectively. In both tunnel diodes left and right   
cold metal electrodes are the same and thus  we obtain anti-symmetric $I$-$V$ curves. 
In both cases with an applied bias voltage $V$ the current $I$ exponentially increases 
and reaches the peak value $I_p$ at about 0.15\,$V$ and thus the first PDR region in both 
tunnel diodes is limited to a small bias voltage window of 0.15\,$V$ in contrast to rough 
estimation of 0.48\,$V$ and 0.6\,$V$ based on the  energy-band diagram in preceding 
section. However, in both tunnel diodes the valley current $I_v$ is determined by 
$W_{\mathrm{m}}$ of electrode materials, i.e., 0.95\,eV for AlI$_2$ and 1.2\,eV for NbS$_2$
(see Table\,\ref{table}), which gives rise to an extended NDR region. At room temperature 
the valley current in lateral tunnel diode is almost zero, which leads to an ultra-high PVCR 
value of 10$^{16}$, while for the vertical tunnel diode the PVCR value is about 10$^{4}$, which 
is still much larger than the conventional semiconductor NDR tunnel diodes. The vertical tunnel 
diode exhibits N-type NDR effect since the electrode material NbS$_2$ possesses  $W_{\mathrm{m}} 
\sim E_{\mathrm{g}}^{\mathrm{V}} \sim E_{\mathrm{g}}^{\mathrm{C}}$ within the PBE-1/2 method 
and thus the second PDR region starts at about 1.2\,$V$  and current increases exponentially 
as in the first PDR region. On the other hand, as will be discussed in detail in the 
following for the lateral tunnel diode the $W_{\mathrm{m}}$ is much smaller than the 
$E_{\mathrm{g}}^{\mathrm{V}}$ and $E_{\mathrm{g}}^{\mathrm{C}}$ (see Figure\,\ref{fig6}a) 
and thus the second PDR region starts at bias voltages larger than 2.3\,$V$. Therefore, the 
NDR effect in the lateral  tunnel diode can be regarded as a $\Lambda$-type NDR.

As expected for both tunnel diodes the current decays exponentially with increasing the 
tunnel barrier thickness as seen in Figure\,\ref{fig5}. For instance, the lateral  
AlI$_2$/MgI$_2$/AlI$_2$ tunnel diode having zigzag orientation posses slightly thicker 
tunnel barrier of about 8.4\,\AA, which leads to 40\% less peak current density compared 
to the tunnel diode with armchair orientation. The situation is very similar in the case
of vertical tunnel diode NbS$_2$/h-BN/NbS$_2$, in which the current decreases by a factor
of 50 from 1 to 2 monolayers of h-BN tunnel barrier. However, the current densities between
the  lateral and vertical tunnel diodes are very different. For a similar tunnel barrier 
thicknesses and the same armchair orientations the lateral tunnel diode has almost four times
larger peak current density of about 290\,$\mu A/\mu m$.

The NDR effect can be explained on the basis of the device density of states (DDOS) (or local 
DOS). For simplicity we focus only on the lateral tunnel diode and consider the armchair 
orientation. The calculated DDOS using the PBE-1/2 method for two different bias voltages are 
presented in Figure\,\ref{fig6}a. Note that in DDOS the energy resolved number of states are 
color coded and it  can be regarded as the energy-band diagram of the tunnel diode (see 
Figure\,\ref{fig3}). As seen from the DDOS and also from the Table\,\ref{table} the band width
$W_{\mathrm{m}}$ of the AlI$_2$ electrode does not change much but $E_{\mathrm{g}}^{\mathrm{V}}$ 
and $E_{\mathrm{g}}^{\mathrm{C}}$ values increases  by 1.3\,eV and 0.8\,eV, respectively within 
the PBE-1/2 method. The tunnel barrier MgI$_2$ has a band  gap of about 5.45\,eV, which is 
clearly seen in DDOS. Moreover, the voltage drop across the tunnel barrier is also seen, 
especially around -3\,eV,  as a linear change of DDOS (the valence band maximum of MgI$_2$) 
from the right to the left electrode. To get further insight into the drop of bias voltage 
across the tunnel barrier in Figure\,\ref{fig6}b we present the electrostatic difference 
potential $\Delta V_{\textrm{E}}$ along the transport direction of the tunnel diode for two 
different bias voltages. As seen the $\Delta V_{\textrm{E}}$ is flat towards the electrodes 
and it drops linearly across the insulating tunnel barrier revealing that the bias voltage 
drop takes place at the tunnel barrier. The potential values are given in Volt and the 
left-to-right potential drop is 0.4\,$V$ and 1.2\,$V$, which are exactly the applied bias 
voltages. Returning back to the discussion of $I$-$V$ characteristics, under an applied bias 
voltage $V$ the current stems from the intra-band tunneling, i.e., electrons in the occupied 
states (see Figure\,\ref{fig4}a) in the right cold metal electrode  can tunnel into the 
unoccupied states of the left cold metal electrode (see also panel 2 in Figure\,\ref{fig3}b). 
This tunneling is very efficient at low bias voltages and thus the peak current turns out 
to be around 0.15\,$V$ and bias voltages higher than the $W_{\mathrm{m}}$ of AlI$_2$ the 
intra-band tunneling completely ceases since occupied and unoccupied energy levels are no 
longer aligned (see panel 4 in Figure\,\ref{fig3}b). This can also be seen from the transmission 
spectrum presented in Figure\,\ref{fig6}c, in which transmission is finite for a bias voltage 
of 0.4\,$V$, whereas it is exactly zero for 1.2\,$V$, which is larger than the $W_{\mathrm{m}}$ 
(0.95\,eV) of AlI$_2$. It is worth to note that in $I$-$V$ characteristics of the  
cold metal tunnel diodes the first PDR and NDR regions stem from the  intra-band tunneling, 
while the second PDR region is due to inter band tunneling. (see panels 2, 3, and 5 in 
Figure\,\ref{fig3}b) Note also that finite PVCR values, especially in $\Lambda$-type NDR 
effect originates from the thermal excitations, i.e, very few high energy electrons at the 
tail of the Fermi-Dirac distribution function could contribute to the valley current.

The $I$-$V$ characteristics and the ultra-high PVCR values of the tunnel diodes discussed 
above are based on the coherent tunneling transport i.e., no dissipation mechanism like 
inelastic electron-electron scattering or electron-phonon scattering is taken into account in 
calculations of the $I$-$V$ curves. The latter is more efficient in 2D semiconductors than
3D ones \cite{cheng2020two}, giving rise to lower room-temperature mobility and as a result 
low on currents in field-effect transistors based on 2D semiconducting nanomaterials \cite{afzalian2021ab}.
The electron-phonon scattering as a dissipation mechanism has been implemented in the 
QuantumATK package, however due to enormous computational cost of such calculations, it has been 
only employed to very simple model systems like H$_2$ molecule placed between 1D Au 
chains.\cite{smidstrup2019an} Thus, for the  present tunnel diodes the inclusion of the 
electron-phonon scattering in transport calculations is not feasible at all from the
computational point of view. Therefore, the peak current $I_1$ (see Figure\,\ref{fig3} 
and Figure\,\ref{fig5}) in the $I$-$V$ characteristics as well as the PVCR value of the 
tunnel diodes can be regarded as the upper limit within the coherent tunneling transport 
approach employed in the present work. Inclusion of the inelastic scattering processes 
like electron-phonon interaction is expected to renormalize the $I$-$V$ characteristics, 
i.e., it would decrease (increase) the peak (valley) current and consequently it would 
reduce the PVCR value. In addition to dissipative processes, defects and interaction with
the environment, i.e., the surrounding dielectric materials would also influence the 
$I$-$V$ characteristics of the devices.  However, the overall $I$-$V$ curves of the tunnel diodes 
are not expected to change substantially, i.e., one can obtain either a conventional N-type 
or $\Lambda$-type NDR effect by proper choice of the cold metal electrode materials.

Finally we would like to comment on the transport properties of the cold metal
MX$_2$ (M=Nb, Ta; X= Se, Se) heterojunctions studied in Ref.\,\cite{yin2022negative}
and dwell on the other applications of the cold metals in nanoelectronics. 
The authors of the Ref.\,\cite{yin2022negative} claim to get a promising PVCR value 
of 51 and extremely high peak current density ( $5.1\times 10^4$ $\mu A$/$\mu m$) for 
an edge contact NbSe$_2$/NbS$_2$ heterojunction device. However, we have shown in 
Supporting Information~\cite{supplement} that the NDR effect obtained in this paper does 
not stem from a physical mechanism but from the numerical treatment of the current in the QuantumATK 
package. Cold metals are not only promising materials for NDR tunnel diodes as discussed 
in this work but also they have been proposed to achieve sub-60 mV/dec subthreshold swing 
(SS) in a CMOS transistor when used as the source electrode. The unique electronic band 
structure of the cold metal source electrode in a CMOS transistor filters the transmission 
of high-energy electrons in the subthreshold region leading to sub-60 mV/dec SS value 
and also reduces the leakage current in the off state
\cite{qiu2018dirac,liu2020switching,marin2020lateral,logoteta2020cold,tang2021steep}.
Besides applications of cold metals in CMOS transistors, in metal-semiconductor 
Schottky barrier diodes the use of a cold metal electrode instead of normal metal 
one breaks the thermionic limit of an ideality factor ($\eta$) of 1 at room temperature, which 
paves the way for the development of low-power electronic circuits \cite{shin2022steep}.

\section{Conclusions}

In conclusion, we proposed a novel semiconductor-free NDR tunnel diode concept that exhibits 
ultra-high PVCR value. The proposed NDR diode consists of two cold metal electrodes separated 
by a thin insulating tunnel barrier. The operation principle and the NDR effect rely on the 
unique electronic band structure of the cold metal electrodes, i.e., the width of the isolated 
metallic bands around the Fermi level as well as the energy gaps separating higher and lower 
lying bands determine the $I$-$V$ characteristics and the PVCR  value of the tunnel diode. We 
showed that by proper choice of the cold metal electrode materials either conventional N-type 
or $\Lambda$-type NDR effect can be obtained. In both types of NDR effect the first PDR region 
and the NDR region are determined by the intra-band quantum tunneling, while inter-band quantum 
tunneling is responsible for the second PDR region in the $I$-$V$ curves of the N-type NDR 
effect. 2D transition-metal dichalcogenides (NbX$_2$, TaX$_2$; X=S, Se, Te), transition-metal 
dihalides (ScI$_2$, YI$_2$, etc), as well as several other 2D nanomaterials possess cold 
metallic electronic band structure and offer a unique platform for the realization of the 
proposed NDR tunnel diodes. We demonstrated the proof of concept by employing the DFT+NEGF 
method and calculated the $I$-$V$ characteristic of the lateral AlI$_2$/MgI$_2$/AlI$_2$ and 
vertical NbS$_2$/h-BN/NbS$_2$ heterojunction tunnel diodes.  For the lateral tunnel diode 
we obtain a $\Lambda$-type NDR effect with an ultra-high PVCR value of 10$^{16}$ at room 
temperature, while the vertical tunnel diode exhibits conventional N-type NDR effect with 
a smaller PVCR value of about 10$^{4}$.  The obtained ultra-high PVCR values in  tunnel 
diodes should be regarded as the upper limit as we neglect the dissipative processes like 
electron-phonon scattering  as well as interaction with environment i.e., the surrounding 
dielectric materials and defects in DFT+NEGF calculations. We anticipate that a plethora of 
2D nanomaterials can serve as cold metal electrodes in proposed tunnel diodes to achieve the 
desired $I$-$V$ characteristics and ultra-high PVCR values in a semiconductor-free solution 
for NDR devices for high-speed, low-power memory and logic applications.

\acknowledgements
E.\c{S} and I.M. acknowledge support from \textit{Sonderforschungsbereich} TRR 227 of 
Deutsche Forschungsgemeinschaft (DFG) and funding provided by the European Union
(EFRE), Grant No: ZS/2016/06/79307.

	
	
%

\bibliography{bibliography}

\begin{thebibliography}{71}
\expandafter\ifx\csname natexlab\endcsname\relax\def\natexlab#1{#1}\fi
\expandafter\ifx\csname bibnamefont\endcsname\relax
  \def\bibnamefont#1{#1}\fi
\expandafter\ifx\csname bibfnamefont\endcsname\relax
  \def\bibfnamefont#1{#1}\fi
\expandafter\ifx\csname citenamefont\endcsname\relax
  \def\citenamefont#1{#1}\fi
\expandafter\ifx\csname url\endcsname\relax
  \def\url#1{\texttt{#1}}\fi
\expandafter\ifx\csname urlprefix\endcsname\relax\def\urlprefix{URL }\fi
\providecommand{\bibinfo}[2]{#2}
\providecommand{\eprint}[2][]{\url{#2}}

\bibitem[{\citenamefont{Moore}(1965)}]{Moore}
\bibinfo{author}{\bibfnamefont{G.~E.} \bibnamefont{Moore}},
  \bibinfo{journal}{Electronics} \textbf{\bibinfo{volume}{38}},
  \bibinfo{pages}{114} (\bibinfo{year}{1965}).

\bibitem[{IRD(2021)}]{IRDS}
\emph{\bibinfo{title}{{International Roadmap for Devices and Systems (IRDS)}}}
  (\bibinfo{year}{2021}), \bibinfo{note}{available at https://irds.ieee.org},
  \urlprefix\url{https://irds.ieee.org/editions/2021}.

\bibitem[{\citenamefont{Chen et~al.}(2014)\citenamefont{Chen, Hutchby, Zhirnov,
  and Bourianoff}}]{chen2014emerging}
\bibinfo{author}{\bibfnamefont{A.}~\bibnamefont{Chen}},
  \bibinfo{author}{\bibfnamefont{J.}~\bibnamefont{Hutchby}},
  \bibinfo{author}{\bibfnamefont{V.}~\bibnamefont{Zhirnov}}, \bibnamefont{and}
  \bibinfo{author}{\bibfnamefont{G.}~\bibnamefont{Bourianoff}},
  \emph{\bibinfo{title}{Emerging Nanoelectronic Devices}}
  (\bibinfo{publisher}{John Wiley \& Sons}, \bibinfo{year}{2014}).

\bibitem[{\citenamefont{Ionescu and Riel}(2011)}]{ionescu2011tunnel}
\bibinfo{author}{\bibfnamefont{A.~M.} \bibnamefont{Ionescu}} \bibnamefont{and}
  \bibinfo{author}{\bibfnamefont{H.}~\bibnamefont{Riel}},
  \bibinfo{journal}{Nature} \textbf{\bibinfo{volume}{479}},
  \bibinfo{pages}{329} (\bibinfo{year}{2011}).

\bibitem[{\citenamefont{Nikonov et~al.}(2011)\citenamefont{Nikonov, Bourianoff,
  and Ghani}}]{nikonov2011proposal}
\bibinfo{author}{\bibfnamefont{D.~E.} \bibnamefont{Nikonov}},
  \bibinfo{author}{\bibfnamefont{G.~I.} \bibnamefont{Bourianoff}},
  \bibnamefont{and} \bibinfo{author}{\bibfnamefont{T.}~\bibnamefont{Ghani}},
  \bibinfo{journal}{IEEE Electron Device Lett.} \textbf{\bibinfo{volume}{32}},
  \bibinfo{pages}{1128} (\bibinfo{year}{2011}).

\bibitem[{\citenamefont{Amlani et~al.}(1999)\citenamefont{Amlani, Orlov, Toth,
  Bernstein, Lent, and Snider}}]{amlani1999digital}
\bibinfo{author}{\bibfnamefont{I.}~\bibnamefont{Amlani}},
  \bibinfo{author}{\bibfnamefont{A.~O.} \bibnamefont{Orlov}},
  \bibinfo{author}{\bibfnamefont{G.}~\bibnamefont{Toth}},
  \bibinfo{author}{\bibfnamefont{G.~H.} \bibnamefont{Bernstein}},
  \bibinfo{author}{\bibfnamefont{C.~S.} \bibnamefont{Lent}}, \bibnamefont{and}
  \bibinfo{author}{\bibfnamefont{G.~L.} \bibnamefont{Snider}},
  \bibinfo{journal}{Science} \textbf{\bibinfo{volume}{284}},
  \bibinfo{pages}{289} (\bibinfo{year}{1999}).

\bibitem[{\citenamefont{Allwood et~al.}(2005)\citenamefont{Allwood, Xiong,
  Faulkner, Atkinson, Petit, and Cowburn}}]{allwood2005magnetic}
\bibinfo{author}{\bibfnamefont{D.~A.} \bibnamefont{Allwood}},
  \bibinfo{author}{\bibfnamefont{G.}~\bibnamefont{Xiong}},
  \bibinfo{author}{\bibfnamefont{C.~C.} \bibnamefont{Faulkner}},
  \bibinfo{author}{\bibfnamefont{D.}~\bibnamefont{Atkinson}},
  \bibinfo{author}{\bibfnamefont{D.}~\bibnamefont{Petit}}, \bibnamefont{and}
  \bibinfo{author}{\bibfnamefont{R.~P.} \bibnamefont{Cowburn}},
  \bibinfo{journal}{Science} \textbf{\bibinfo{volume}{309}},
  \bibinfo{pages}{1688} (\bibinfo{year}{2005}).

\bibitem[{\citenamefont{Pan and Naeemi}(2017)}]{pan2017expanded}
\bibinfo{author}{\bibfnamefont{C.}~\bibnamefont{Pan}} \bibnamefont{and}
  \bibinfo{author}{\bibfnamefont{A.}~\bibnamefont{Naeemi}},
  \bibinfo{journal}{IEEE J. Explor. Solid-State Comput. Devices Circuits}
  \textbf{\bibinfo{volume}{3}}, \bibinfo{pages}{101} (\bibinfo{year}{2017}).

\bibitem[{\citenamefont{Berger and Ramesh}(2011)}]{berger2011negative}
\bibinfo{author}{\bibfnamefont{P.~R.} \bibnamefont{Berger}} \bibnamefont{and}
  \bibinfo{author}{\bibfnamefont{A.}~\bibnamefont{Ramesh}}, in
  \emph{\bibinfo{booktitle}{Comprehensive Semiconductor Science and
  Technology}} (\bibinfo{publisher}{Elsevier Inc.}, \bibinfo{year}{2011}), pp.
  \bibinfo{pages}{176--241}.

\bibitem[{\citenamefont{Jo et~al.}(2021)\citenamefont{Jo, Kang, and
  Cho}}]{jo2021recent}
\bibinfo{author}{\bibfnamefont{S.~B.} \bibnamefont{Jo}},
  \bibinfo{author}{\bibfnamefont{J.}~\bibnamefont{Kang}}, \bibnamefont{and}
  \bibinfo{author}{\bibfnamefont{J.~H.} \bibnamefont{Cho}},
  \bibinfo{journal}{Advanced Science} \textbf{\bibinfo{volume}{8}},
  \bibinfo{pages}{2004216} (\bibinfo{year}{2021}).

\bibitem[{\citenamefont{Van Der~Wagt}(1999)}]{van1999tunneling}
\bibinfo{author}{\bibfnamefont{J.}~\bibnamefont{Van Der~Wagt}},
  \bibinfo{journal}{Proceedings of the IEEE} \textbf{\bibinfo{volume}{87}},
  \bibinfo{pages}{571} (\bibinfo{year}{1999}).

\bibitem[{\citenamefont{Karda et~al.}(2009)\citenamefont{Karda, Brockman,
  Sutar, Seabaugh, and Nahas}}]{karda2009one}
\bibinfo{author}{\bibfnamefont{K.}~\bibnamefont{Karda}},
  \bibinfo{author}{\bibfnamefont{J.}~\bibnamefont{Brockman}},
  \bibinfo{author}{\bibfnamefont{S.}~\bibnamefont{Sutar}},
  \bibinfo{author}{\bibfnamefont{A.}~\bibnamefont{Seabaugh}}, \bibnamefont{and}
  \bibinfo{author}{\bibfnamefont{J.}~\bibnamefont{Nahas}}, in
  \emph{\bibinfo{booktitle}{2009 IEEE International Conference on IC Design and
  Technology}} (\bibinfo{organization}{IEEE}, \bibinfo{year}{2009}), pp.
  \bibinfo{pages}{233--236}.

\bibitem[{\citenamefont{Chen et~al.}(1996)\citenamefont{Chen, Maezawa, and
  Yamamoto}}]{chen1996inp}
\bibinfo{author}{\bibfnamefont{K.~J.} \bibnamefont{Chen}},
  \bibinfo{author}{\bibfnamefont{K.}~\bibnamefont{Maezawa}}, \bibnamefont{and}
  \bibinfo{author}{\bibfnamefont{M.}~\bibnamefont{Yamamoto}},
  \bibinfo{journal}{IEEE Electron Device Lett.} \textbf{\bibinfo{volume}{17}},
  \bibinfo{pages}{127} (\bibinfo{year}{1996}).

\bibitem[{\citenamefont{Maezawa et~al.}(1998)\citenamefont{Maezawa, Matsuzaki,
  Yamamoto, and Otsuji}}]{maezawa1998high}
\bibinfo{author}{\bibfnamefont{K.}~\bibnamefont{Maezawa}},
  \bibinfo{author}{\bibfnamefont{H.}~\bibnamefont{Matsuzaki}},
  \bibinfo{author}{\bibfnamefont{M.}~\bibnamefont{Yamamoto}}, \bibnamefont{and}
  \bibinfo{author}{\bibfnamefont{T.}~\bibnamefont{Otsuji}},
  \bibinfo{journal}{IEEE Electron Device Lett.} \textbf{\bibinfo{volume}{19}},
  \bibinfo{pages}{80} (\bibinfo{year}{1998}).

\bibitem[{\citenamefont{Williamson et~al.}(1997)\citenamefont{Williamson,
  Enquist, Chow, Dunlap, Subramaniam, Lei, Bernstein, and
  Gilbert}}]{williamson199712}
\bibinfo{author}{\bibfnamefont{W.}~\bibnamefont{Williamson}},
  \bibinfo{author}{\bibfnamefont{S.~B.} \bibnamefont{Enquist}},
  \bibinfo{author}{\bibfnamefont{D.~H.} \bibnamefont{Chow}},
  \bibinfo{author}{\bibfnamefont{H.~L.} \bibnamefont{Dunlap}},
  \bibinfo{author}{\bibfnamefont{S.}~\bibnamefont{Subramaniam}},
  \bibinfo{author}{\bibfnamefont{P.}~\bibnamefont{Lei}},
  \bibinfo{author}{\bibfnamefont{G.~H.} \bibnamefont{Bernstein}},
  \bibnamefont{and} \bibinfo{author}{\bibfnamefont{B.~K.}
  \bibnamefont{Gilbert}}, \bibinfo{journal}{IEEE J. Solid-State Circuits}
  \textbf{\bibinfo{volume}{32}}, \bibinfo{pages}{222} (\bibinfo{year}{1997}).

\bibitem[{\citenamefont{Micheel and Paulus}(1990)}]{micheel1990differential}
\bibinfo{author}{\bibfnamefont{L.~J.} \bibnamefont{Micheel}} \bibnamefont{and}
  \bibinfo{author}{\bibfnamefont{M.~J.} \bibnamefont{Paulus}}, in
  \emph{\bibinfo{booktitle}{Proceedings of the Twentieth International
  Symposium on Multiple-Valued Logic}} (\bibinfo{organization}{IEEE Computer
  Society}, \bibinfo{year}{1990}), pp. \bibinfo{pages}{189--190}.

\bibitem[{\citenamefont{Lin}(1994)}]{lin1994resonant}
\bibinfo{author}{\bibfnamefont{H.}~\bibnamefont{Lin}}, in
  \emph{\bibinfo{booktitle}{Proceedings of 24th International Symposium on
  Multiple-Valued Logic (ISMVL'94)}} (\bibinfo{organization}{IEEE},
  \bibinfo{year}{1994}), pp. \bibinfo{pages}{188--195}.

\bibitem[{\citenamefont{Jin et~al.}(2004)\citenamefont{Jin, Chung, Heyns,
  Berger, Yu, Thompson, and Rommel}}]{jin2004tri}
\bibinfo{author}{\bibfnamefont{N.}~\bibnamefont{Jin}},
  \bibinfo{author}{\bibfnamefont{S.-Y.} \bibnamefont{Chung}},
  \bibinfo{author}{\bibfnamefont{R.~M.} \bibnamefont{Heyns}},
  \bibinfo{author}{\bibfnamefont{P.~R.} \bibnamefont{Berger}},
  \bibinfo{author}{\bibfnamefont{R.}~\bibnamefont{Yu}},
  \bibinfo{author}{\bibfnamefont{P.~E.} \bibnamefont{Thompson}},
  \bibnamefont{and} \bibinfo{author}{\bibfnamefont{S.~L.}
  \bibnamefont{Rommel}}, \bibinfo{journal}{IEEE Electron Device Lett.}
  \textbf{\bibinfo{volume}{25}}, \bibinfo{pages}{646} (\bibinfo{year}{2004}).

\bibitem[{\citenamefont{Hurst}(1984)}]{hurst1984multiple}
\bibinfo{author}{\bibfnamefont{S.~L.} \bibnamefont{Hurst}},
  \bibinfo{journal}{IEEE Trans. Comput.} \textbf{\bibinfo{volume}{33}},
  \bibinfo{pages}{1160} (\bibinfo{year}{1984}).

\bibitem[{\citenamefont{Esaki}(1958)}]{esaki1958new}
\bibinfo{author}{\bibfnamefont{L.}~\bibnamefont{Esaki}},
  \bibinfo{journal}{Phys. Rev.} \textbf{\bibinfo{volume}{109}},
  \bibinfo{pages}{603} (\bibinfo{year}{1958}).

\bibitem[{\citenamefont{Esaki and Tsu}(1970)}]{esaki1970superlattice}
\bibinfo{author}{\bibfnamefont{L.}~\bibnamefont{Esaki}} \bibnamefont{and}
  \bibinfo{author}{\bibfnamefont{R.}~\bibnamefont{Tsu}}, \bibinfo{journal}{IBM
  J. Res. Dev.} \textbf{\bibinfo{volume}{14}}, \bibinfo{pages}{61}
  (\bibinfo{year}{1970}).

\bibitem[{\citenamefont{Ramesh et~al.}(2012)\citenamefont{Ramesh, Berger, and
  Loo}}]{ramesh2012high}
\bibinfo{author}{\bibfnamefont{A.}~\bibnamefont{Ramesh}},
  \bibinfo{author}{\bibfnamefont{P.~R.} \bibnamefont{Berger}},
  \bibnamefont{and} \bibinfo{author}{\bibfnamefont{R.}~\bibnamefont{Loo}},
  \bibinfo{journal}{Appl. Phys. Lett.} \textbf{\bibinfo{volume}{100}},
  \bibinfo{pages}{092104} (\bibinfo{year}{2012}).

\bibitem[{\citenamefont{Fung et~al.}(2011)\citenamefont{Fung, Chen, and
  Lu}}]{fung2011esaki}
\bibinfo{author}{\bibfnamefont{W.~Y.} \bibnamefont{Fung}},
  \bibinfo{author}{\bibfnamefont{L.}~\bibnamefont{Chen}}, \bibnamefont{and}
  \bibinfo{author}{\bibfnamefont{W.}~\bibnamefont{Lu}}, \bibinfo{journal}{Appl.
  Phys. Lett.} \textbf{\bibinfo{volume}{99}}, \bibinfo{pages}{092108}
  (\bibinfo{year}{2011}).

\bibitem[{\citenamefont{Duschl et~al.}(1999)\citenamefont{Duschl, Schmidt,
  Reitemann, Kasper, and Eberl}}]{duschl1999high}
\bibinfo{author}{\bibfnamefont{R.}~\bibnamefont{Duschl}},
  \bibinfo{author}{\bibfnamefont{O.~G.} \bibnamefont{Schmidt}},
  \bibinfo{author}{\bibfnamefont{G.}~\bibnamefont{Reitemann}},
  \bibinfo{author}{\bibfnamefont{E.}~\bibnamefont{Kasper}}, \bibnamefont{and}
  \bibinfo{author}{\bibfnamefont{K.}~\bibnamefont{Eberl}},
  \bibinfo{journal}{Electron. Lett.} \textbf{\bibinfo{volume}{35}},
  \bibinfo{pages}{1111} (\bibinfo{year}{1999}).

\bibitem[{\citenamefont{Schmid et~al.}(2012)\citenamefont{Schmid, Bessire,
  Bj\"ork, Schenk, and Riel}}]{schmid2012silicon}
\bibinfo{author}{\bibfnamefont{H.}~\bibnamefont{Schmid}},
  \bibinfo{author}{\bibfnamefont{C.}~\bibnamefont{Bessire}},
  \bibinfo{author}{\bibfnamefont{M.~T.} \bibnamefont{Bj\"ork}},
  \bibinfo{author}{\bibfnamefont{A.}~\bibnamefont{Schenk}}, \bibnamefont{and}
  \bibinfo{author}{\bibfnamefont{H.}~\bibnamefont{Riel}},
  \bibinfo{journal}{Nano Lett.} \textbf{\bibinfo{volume}{12}},
  \bibinfo{pages}{699} (\bibinfo{year}{2012}).

\bibitem[{\citenamefont{Chow et~al.}(1992)\citenamefont{Chow, Schulman,
  {\"O}zbay, and Bloom}}]{chow1992investigation}
\bibinfo{author}{\bibfnamefont{D.~H.} \bibnamefont{Chow}},
  \bibinfo{author}{\bibfnamefont{J.~N.} \bibnamefont{Schulman}},
  \bibinfo{author}{\bibfnamefont{E.}~\bibnamefont{{\"O}zbay}},
  \bibnamefont{and} \bibinfo{author}{\bibfnamefont{D.~M.} \bibnamefont{Bloom}},
  \bibinfo{journal}{Appl. Phys. Lett.} \textbf{\bibinfo{volume}{61}},
  \bibinfo{pages}{1685} (\bibinfo{year}{1992}).

\bibitem[{\citenamefont{Tsai et~al.}(1994)\citenamefont{Tsai, Su, Lin, Wang,
  and Lee}}]{tsai1994pn}
\bibinfo{author}{\bibfnamefont{H.}~\bibnamefont{Tsai}},
  \bibinfo{author}{\bibfnamefont{Y.}~\bibnamefont{Su}},
  \bibinfo{author}{\bibfnamefont{H.}~\bibnamefont{Lin}},
  \bibinfo{author}{\bibfnamefont{R.}~\bibnamefont{Wang}}, \bibnamefont{and}
  \bibinfo{author}{\bibfnamefont{T.}~\bibnamefont{Lee}}, \bibinfo{journal}{IEEE
  Electron Device Lett.} \textbf{\bibinfo{volume}{15}}, \bibinfo{pages}{357}
  (\bibinfo{year}{1994}).

\bibitem[{\citenamefont{Sant and Schenk}(2017)}]{sant2017effect}
\bibinfo{author}{\bibfnamefont{S.}~\bibnamefont{Sant}} \bibnamefont{and}
  \bibinfo{author}{\bibfnamefont{A.}~\bibnamefont{Schenk}},
  \bibinfo{journal}{J. Appl. Phys.} \textbf{\bibinfo{volume}{122}},
  \bibinfo{pages}{135702} (\bibinfo{year}{2017}).

\bibitem[{\citenamefont{Bizindavyi et~al.}(2018)\citenamefont{Bizindavyi,
  Verhulst, Smets, Verreck, Sor{\'e}e, and Groeseneken}}]{bizindavyi2018band}
\bibinfo{author}{\bibfnamefont{J.}~\bibnamefont{Bizindavyi}},
  \bibinfo{author}{\bibfnamefont{A.~S.} \bibnamefont{Verhulst}},
  \bibinfo{author}{\bibfnamefont{Q.}~\bibnamefont{Smets}},
  \bibinfo{author}{\bibfnamefont{D.}~\bibnamefont{Verreck}},
  \bibinfo{author}{\bibfnamefont{B.}~\bibnamefont{Sor{\'e}e}},
  \bibnamefont{and}
  \bibinfo{author}{\bibfnamefont{G.}~\bibnamefont{Groeseneken}},
  \bibinfo{journal}{IEEE J. Electron Devices Soc.}
  \textbf{\bibinfo{volume}{6}}, \bibinfo{pages}{633} (\bibinfo{year}{2018}).

\bibitem[{\citenamefont{Schenk and Sant}(2020)}]{schenk2020tunneling}
\bibinfo{author}{\bibfnamefont{A.}~\bibnamefont{Schenk}} \bibnamefont{and}
  \bibinfo{author}{\bibfnamefont{S.}~\bibnamefont{Sant}}, \bibinfo{journal}{J.
  Appl. Phys.} \textbf{\bibinfo{volume}{128}}, \bibinfo{pages}{014502}
  (\bibinfo{year}{2020}).

\bibitem[{\citenamefont{Chung et~al.}(2004)\citenamefont{Chung, Jin, Berger,
  Yu, Thompson, Lake, Rommel, and Kurinec}}]{chung2004three}
\bibinfo{author}{\bibfnamefont{S.-Y.} \bibnamefont{Chung}},
  \bibinfo{author}{\bibfnamefont{N.}~\bibnamefont{Jin}},
  \bibinfo{author}{\bibfnamefont{P.~R.} \bibnamefont{Berger}},
  \bibinfo{author}{\bibfnamefont{R.}~\bibnamefont{Yu}},
  \bibinfo{author}{\bibfnamefont{P.~E.} \bibnamefont{Thompson}},
  \bibinfo{author}{\bibfnamefont{R.}~\bibnamefont{Lake}},
  \bibinfo{author}{\bibfnamefont{S.~L.} \bibnamefont{Rommel}},
  \bibnamefont{and} \bibinfo{author}{\bibfnamefont{S.~K.}
  \bibnamefont{Kurinec}}, \bibinfo{journal}{Appl. Phys. Lett.}
  \textbf{\bibinfo{volume}{84}}, \bibinfo{pages}{2688} (\bibinfo{year}{2004}).

\bibitem[{\citenamefont{Chen et~al.}(2009)\citenamefont{Chen, Griffin, and
  Plummer}}]{chen2009negative}
\bibinfo{author}{\bibfnamefont{S.-L.} \bibnamefont{Chen}},
  \bibinfo{author}{\bibfnamefont{P.~B.} \bibnamefont{Griffin}},
  \bibnamefont{and} \bibinfo{author}{\bibfnamefont{J.~D.}
  \bibnamefont{Plummer}}, \bibinfo{journal}{IEEE Trans. Electron Devices}
  \textbf{\bibinfo{volume}{56}}, \bibinfo{pages}{634} (\bibinfo{year}{2009}).

\bibitem[{\citenamefont{Gan et~al.}(2007)\citenamefont{Gan, Tsai, and
  Sun}}]{gan2007fabrication}
\bibinfo{author}{\bibfnamefont{K.-J.} \bibnamefont{Gan}},
  \bibinfo{author}{\bibfnamefont{C.-S.} \bibnamefont{Tsai}}, \bibnamefont{and}
  \bibinfo{author}{\bibfnamefont{W.-L.} \bibnamefont{Sun}},
  \bibinfo{journal}{Electron. Lett.} \textbf{\bibinfo{volume}{43}},
  \bibinfo{pages}{517} (\bibinfo{year}{2007}).

\bibitem[{\citenamefont{Duane et~al.}(2003)\citenamefont{Duane, Mathewson, and
  Concannon}}]{duane2003bistable}
\bibinfo{author}{\bibfnamefont{R.}~\bibnamefont{Duane}},
  \bibinfo{author}{\bibfnamefont{A.}~\bibnamefont{Mathewson}},
  \bibnamefont{and}
  \bibinfo{author}{\bibfnamefont{A.}~\bibnamefont{Concannon}},
  \bibinfo{journal}{IEEE Electron Device Lett.} \textbf{\bibinfo{volume}{24}},
  \bibinfo{pages}{661} (\bibinfo{year}{2003}).

\bibitem[{\citenamefont{Fang et~al.}(2022)\citenamefont{Fang, Qiu, Zhang, Hu,
  and Peng}}]{fang2022giant}
\bibinfo{author}{\bibfnamefont{L.}~\bibnamefont{Fang}},
  \bibinfo{author}{\bibfnamefont{C.}~\bibnamefont{Qiu}},
  \bibinfo{author}{\bibfnamefont{H.}~\bibnamefont{Zhang}},
  \bibinfo{author}{\bibfnamefont{Y.}~\bibnamefont{Hu}}, \bibnamefont{and}
  \bibinfo{author}{\bibfnamefont{L.-M.} \bibnamefont{Peng}},
  \bibinfo{journal}{Adv. Electron. Mater.} p. \bibinfo{pages}{2101314}
  (\bibinfo{year}{2022}).

\bibitem[{\citenamefont{Gan et~al.}(2016)\citenamefont{Gan, Lu, Yeh, Chen, and
  Chen}}]{gan2016multiple}
\bibinfo{author}{\bibfnamefont{K.-J.} \bibnamefont{Gan}},
  \bibinfo{author}{\bibfnamefont{J.-J.} \bibnamefont{Lu}},
  \bibinfo{author}{\bibfnamefont{W.-K.} \bibnamefont{Yeh}},
  \bibinfo{author}{\bibfnamefont{Y.-H.} \bibnamefont{Chen}}, \bibnamefont{and}
  \bibinfo{author}{\bibfnamefont{Y.-W.} \bibnamefont{Chen}},
  \bibinfo{journal}{Eng. Sci. Technol. an Int. J.}
  \textbf{\bibinfo{volume}{19}}, \bibinfo{pages}{888} (\bibinfo{year}{2016}).

\bibitem[{\citenamefont{{\c{S}}a{\c{s}}{\i}o{\u{g}}lu{ } and
  Mertig}(2021)}]{sasioglu2021patent}
\bibinfo{author}{\bibfnamefont{E.}~\bibnamefont{{\c{S}}a{\c{s}}{\i}o{\u{g}}lu{
  }}} \bibnamefont{and}
  \bibinfo{author}{\bibfnamefont{I.}~\bibnamefont{Mertig}},
  \emph{\bibinfo{title}{{Negative diﬀerential resistance tunnel diode and
  manufacturing method}}} (\bibinfo{year}{2021}), \bibinfo{note}{{German Patent
  Application 102021206526.0}}.

\bibitem[{\citenamefont{Sasioglu and Mertig}(2022)}]{sasioglu2022negative}
\bibinfo{author}{\bibfnamefont{E.}~\bibnamefont{Sasioglu}} \bibnamefont{and}
  \bibinfo{author}{\bibfnamefont{I.}~\bibnamefont{Mertig}},
  \bibinfo{journal}{Bull. Am. Phys. Soc.}  (\bibinfo{year}{2022}).

\bibitem[{\citenamefont{Bardeen}(1961)}]{bardeen1961tunnelling}
\bibinfo{author}{\bibfnamefont{J.}~\bibnamefont{Bardeen}},
  \bibinfo{journal}{Phys. Rev. Lett.} \textbf{\bibinfo{volume}{6}},
  \bibinfo{pages}{57} (\bibinfo{year}{1961}).

\bibitem[{\citenamefont{Meservey and Tedrow}(1994)}]{meservey1994spin}
\bibinfo{author}{\bibfnamefont{R.}~\bibnamefont{Meservey}} \bibnamefont{and}
  \bibinfo{author}{\bibfnamefont{P.}~\bibnamefont{Tedrow}},
  \bibinfo{journal}{Phys. Rep.} \textbf{\bibinfo{volume}{238}},
  \bibinfo{pages}{173} (\bibinfo{year}{1994}).

\bibitem[{\citenamefont{Gunnarsson}(2004)}]{gunnarsson2004alkali}
\bibinfo{author}{\bibfnamefont{O.}~\bibnamefont{Gunnarsson}},
  \emph{\bibinfo{title}{Alkali-doped fullerides: narrow-band solids with
  unusual properties}} (\bibinfo{publisher}{World Scientific},
  \bibinfo{year}{2004}).

\bibitem[{\citenamefont{Hansmann et~al.}(2013)\citenamefont{Hansmann, Toschi,
  Sangiovanni, Saha-Dasgupta, Lupi, Marsi, and Held}}]{hansmann2013mott}
\bibinfo{author}{\bibfnamefont{P.}~\bibnamefont{Hansmann}},
  \bibinfo{author}{\bibfnamefont{A.}~\bibnamefont{Toschi}},
  \bibinfo{author}{\bibfnamefont{G.}~\bibnamefont{Sangiovanni}},
  \bibinfo{author}{\bibfnamefont{T.}~\bibnamefont{Saha-Dasgupta}},
  \bibinfo{author}{\bibfnamefont{S.}~\bibnamefont{Lupi}},
  \bibinfo{author}{\bibfnamefont{M.}~\bibnamefont{Marsi}}, \bibnamefont{and}
  \bibinfo{author}{\bibfnamefont{K.}~\bibnamefont{Held}},
  \emph{\bibinfo{title}{Mott--hubbard transition in v$_2$o$_3$ revisited}}
  (\bibinfo{year}{2013}).

\bibitem[{\citenamefont{Regnault et~al.}(2022)\citenamefont{Regnault, Xu, Li,
  Ma, Jovanovic, Yazdani, Parkin, Felser, Schoop, Ong
  et~al.}}]{regnault2022catalogue}
\bibinfo{author}{\bibfnamefont{N.}~\bibnamefont{Regnault}},
  \bibinfo{author}{\bibfnamefont{Y.}~\bibnamefont{Xu}},
  \bibinfo{author}{\bibfnamefont{M.-R.} \bibnamefont{Li}},
  \bibinfo{author}{\bibfnamefont{D.-S.} \bibnamefont{Ma}},
  \bibinfo{author}{\bibfnamefont{M.}~\bibnamefont{Jovanovic}},
  \bibinfo{author}{\bibfnamefont{A.}~\bibnamefont{Yazdani}},
  \bibinfo{author}{\bibfnamefont{S.~S.~P.} \bibnamefont{Parkin}},
  \bibinfo{author}{\bibfnamefont{C.}~\bibnamefont{Felser}},
  \bibinfo{author}{\bibfnamefont{L.~M.} \bibnamefont{Schoop}},
  \bibinfo{author}{\bibfnamefont{N.~P.} \bibnamefont{Ong}},
  \bibnamefont{et~al.}, \bibinfo{journal}{Nature}
  \textbf{\bibinfo{volume}{603}}, \bibinfo{pages}{824} (\bibinfo{year}{2022}).

\bibitem[{\citenamefont{Haastrup et~al.}(2018)\citenamefont{Haastrup, Strange,
  Pandey, Deilmann, Schmidt, Hinsche, Gjerding, Torelli, Larsen, Riis-Jensen
  et~al.}}]{haastrup2018computational}
\bibinfo{author}{\bibfnamefont{S.}~\bibnamefont{Haastrup}},
  \bibinfo{author}{\bibfnamefont{M.}~\bibnamefont{Strange}},
  \bibinfo{author}{\bibfnamefont{M.}~\bibnamefont{Pandey}},
  \bibinfo{author}{\bibfnamefont{T.}~\bibnamefont{Deilmann}},
  \bibinfo{author}{\bibfnamefont{P.~S.} \bibnamefont{Schmidt}},
  \bibinfo{author}{\bibfnamefont{N.~F.} \bibnamefont{Hinsche}},
  \bibinfo{author}{\bibfnamefont{M.~N.} \bibnamefont{Gjerding}},
  \bibinfo{author}{\bibfnamefont{D.}~\bibnamefont{Torelli}},
  \bibinfo{author}{\bibfnamefont{P.~M.} \bibnamefont{Larsen}},
  \bibinfo{author}{\bibfnamefont{A.~C.} \bibnamefont{Riis-Jensen}},
  \bibnamefont{et~al.}, \bibinfo{journal}{2d Mater.}
  \textbf{\bibinfo{volume}{5}}, \bibinfo{pages}{042002} (\bibinfo{year}{2018}).

\bibitem[{\citenamefont{Gjerding et~al.}(2021)\citenamefont{Gjerding,
  Taghizadeh, Rasmussen, Ali, Bertoldo, Deilmann, Kn{\o}sgaard, Kruse, Larsen,
  Manti et~al.}}]{gjerding2021recent}
\bibinfo{author}{\bibfnamefont{M.~N.} \bibnamefont{Gjerding}},
  \bibinfo{author}{\bibfnamefont{A.}~\bibnamefont{Taghizadeh}},
  \bibinfo{author}{\bibfnamefont{A.}~\bibnamefont{Rasmussen}},
  \bibinfo{author}{\bibfnamefont{S.}~\bibnamefont{Ali}},
  \bibinfo{author}{\bibfnamefont{F.}~\bibnamefont{Bertoldo}},
  \bibinfo{author}{\bibfnamefont{T.}~\bibnamefont{Deilmann}},
  \bibinfo{author}{\bibfnamefont{N.~R.} \bibnamefont{Kn{\o}sgaard}},
  \bibinfo{author}{\bibfnamefont{M.}~\bibnamefont{Kruse}},
  \bibinfo{author}{\bibfnamefont{A.~H.} \bibnamefont{Larsen}},
  \bibinfo{author}{\bibfnamefont{S.}~\bibnamefont{Manti}},
  \bibnamefont{et~al.}, \bibinfo{journal}{2d Mater.}
  \textbf{\bibinfo{volume}{8}}, \bibinfo{pages}{044002} (\bibinfo{year}{2021}).

\bibitem[{\citenamefont{Mounet et~al.}(2018)\citenamefont{Mounet, Gibertini,
  Schwaller, Campi, Merkys, Marrazzo, Sohier, Castelli, Cepellotti, Pizzi
  et~al.}}]{mounet2018two}
\bibinfo{author}{\bibfnamefont{N.}~\bibnamefont{Mounet}},
  \bibinfo{author}{\bibfnamefont{M.}~\bibnamefont{Gibertini}},
  \bibinfo{author}{\bibfnamefont{P.}~\bibnamefont{Schwaller}},
  \bibinfo{author}{\bibfnamefont{D.}~\bibnamefont{Campi}},
  \bibinfo{author}{\bibfnamefont{A.}~\bibnamefont{Merkys}},
  \bibinfo{author}{\bibfnamefont{A.}~\bibnamefont{Marrazzo}},
  \bibinfo{author}{\bibfnamefont{T.}~\bibnamefont{Sohier}},
  \bibinfo{author}{\bibfnamefont{I.~E.} \bibnamefont{Castelli}},
  \bibinfo{author}{\bibfnamefont{A.}~\bibnamefont{Cepellotti}},
  \bibinfo{author}{\bibfnamefont{G.}~\bibnamefont{Pizzi}},
  \bibnamefont{et~al.}, \bibinfo{journal}{Nat. Nanotechnol.}
  \textbf{\bibinfo{volume}{13}}, \bibinfo{pages}{246} (\bibinfo{year}{2018}).

\bibitem[{\citenamefont{Janatipour et~al.}(2022)\citenamefont{Janatipour,
  Mahdavifar, Noorizadeh, and Schreckenbach}}]{janatipour2022achieved}
\bibinfo{author}{\bibfnamefont{N.}~\bibnamefont{Janatipour}},
  \bibinfo{author}{\bibfnamefont{Z.}~\bibnamefont{Mahdavifar}},
  \bibinfo{author}{\bibfnamefont{S.}~\bibnamefont{Noorizadeh}},
  \bibnamefont{and}
  \bibinfo{author}{\bibfnamefont{G.}~\bibnamefont{Schreckenbach}},
  \bibinfo{journal}{RSC Advances} \textbf{\bibinfo{volume}{12}},
  \bibinfo{pages}{1758} (\bibinfo{year}{2022}).

\bibitem[{\citenamefont{Ostovan et~al.}(2018)\citenamefont{Ostovan, Mahdavifar,
  and Bamdad}}]{ostovan2018length}
\bibinfo{author}{\bibfnamefont{A.}~\bibnamefont{Ostovan}},
  \bibinfo{author}{\bibfnamefont{Z.}~\bibnamefont{Mahdavifar}},
  \bibnamefont{and} \bibinfo{author}{\bibfnamefont{M.}~\bibnamefont{Bamdad}},
  \bibinfo{journal}{J. Mol. Liq.} \textbf{\bibinfo{volume}{269}},
  \bibinfo{pages}{639} (\bibinfo{year}{2018}).

\bibitem[{\citenamefont{Mishchenko et~al.}(2014)\citenamefont{Mishchenko, Tu,
  Cao, Gorbachev, Wallbank, Greenaway, Morozov, Morozov, Zhu, Wong
  et~al.}}]{mishchenko2014twist}
\bibinfo{author}{\bibfnamefont{A.}~\bibnamefont{Mishchenko}},
  \bibinfo{author}{\bibfnamefont{J.}~\bibnamefont{Tu}},
  \bibinfo{author}{\bibfnamefont{Y.}~\bibnamefont{Cao}},
  \bibinfo{author}{\bibfnamefont{R.~V.} \bibnamefont{Gorbachev}},
  \bibinfo{author}{\bibfnamefont{J.}~\bibnamefont{Wallbank}},
  \bibinfo{author}{\bibfnamefont{M.~T.} \bibnamefont{Greenaway}},
  \bibinfo{author}{\bibfnamefont{V.}~\bibnamefont{Morozov}},
  \bibinfo{author}{\bibfnamefont{S.}~\bibnamefont{Morozov}},
  \bibinfo{author}{\bibfnamefont{M.}~\bibnamefont{Zhu}},
  \bibinfo{author}{\bibfnamefont{S.}~\bibnamefont{Wong}}, \bibnamefont{et~al.},
  \bibinfo{journal}{Nat. Nanotechnol.} \textbf{\bibinfo{volume}{9}},
  \bibinfo{pages}{808} (\bibinfo{year}{2014}).

\bibitem[{\citenamefont{Shim et~al.}(2016)\citenamefont{Shim, Oh, Kang, Jo,
  Ali, Choi, Heo, Jeon, Lee, Kim et~al.}}]{shim2016phosphorene}
\bibinfo{author}{\bibfnamefont{J.}~\bibnamefont{Shim}},
  \bibinfo{author}{\bibfnamefont{S.}~\bibnamefont{Oh}},
  \bibinfo{author}{\bibfnamefont{D.-H.} \bibnamefont{Kang}},
  \bibinfo{author}{\bibfnamefont{S.-H.} \bibnamefont{Jo}},
  \bibinfo{author}{\bibfnamefont{M.~H.} \bibnamefont{Ali}},
  \bibinfo{author}{\bibfnamefont{W.-Y.} \bibnamefont{Choi}},
  \bibinfo{author}{\bibfnamefont{K.}~\bibnamefont{Heo}},
  \bibinfo{author}{\bibfnamefont{J.}~\bibnamefont{Jeon}},
  \bibinfo{author}{\bibfnamefont{S.}~\bibnamefont{Lee}},
  \bibinfo{author}{\bibfnamefont{M.}~\bibnamefont{Kim}}, \bibnamefont{et~al.},
  \bibinfo{journal}{Nat. Commun.} \textbf{\bibinfo{volume}{7}},
  \bibinfo{pages}{1} (\bibinfo{year}{2016}).

\bibitem[{\citenamefont{Kim et~al.}(2020)\citenamefont{Kim, Park, Shim, Shin,
  Andreev, Koo, Yoo, Jung, Heo, Lee et~al.}}]{kim2020multiple}
\bibinfo{author}{\bibfnamefont{K.-H.} \bibnamefont{Kim}},
  \bibinfo{author}{\bibfnamefont{H.-Y.} \bibnamefont{Park}},
  \bibinfo{author}{\bibfnamefont{J.}~\bibnamefont{Shim}},
  \bibinfo{author}{\bibfnamefont{G.}~\bibnamefont{Shin}},
  \bibinfo{author}{\bibfnamefont{M.}~\bibnamefont{Andreev}},
  \bibinfo{author}{\bibfnamefont{J.}~\bibnamefont{Koo}},
  \bibinfo{author}{\bibfnamefont{G.}~\bibnamefont{Yoo}},
  \bibinfo{author}{\bibfnamefont{K.}~\bibnamefont{Jung}},
  \bibinfo{author}{\bibfnamefont{K.}~\bibnamefont{Heo}},
  \bibinfo{author}{\bibfnamefont{Y.}~\bibnamefont{Lee}}, \bibnamefont{et~al.},
  \bibinfo{journal}{Nanoscale Horiz.} \textbf{\bibinfo{volume}{5}},
  \bibinfo{pages}{654} (\bibinfo{year}{2020}).

\bibitem[{\citenamefont{Lin et~al.}(2015)\citenamefont{Lin, Ghosh, Addou, Lu,
  Eichfeld, Zhu, Li, Peng, Kim, Li et~al.}}]{lin2015atomically}
\bibinfo{author}{\bibfnamefont{Y.-C.} \bibnamefont{Lin}},
  \bibinfo{author}{\bibfnamefont{R.~K.} \bibnamefont{Ghosh}},
  \bibinfo{author}{\bibfnamefont{R.}~\bibnamefont{Addou}},
  \bibinfo{author}{\bibfnamefont{N.}~\bibnamefont{Lu}},
  \bibinfo{author}{\bibfnamefont{S.~M.} \bibnamefont{Eichfeld}},
  \bibinfo{author}{\bibfnamefont{H.}~\bibnamefont{Zhu}},
  \bibinfo{author}{\bibfnamefont{M.-Y.} \bibnamefont{Li}},
  \bibinfo{author}{\bibfnamefont{X.}~\bibnamefont{Peng}},
  \bibinfo{author}{\bibfnamefont{M.~J.} \bibnamefont{Kim}},
  \bibinfo{author}{\bibfnamefont{L.-J.} \bibnamefont{Li}},
  \bibnamefont{et~al.}, \bibinfo{journal}{Nat. Commun.}
  \textbf{\bibinfo{volume}{6}}, \bibinfo{pages}{1} (\bibinfo{year}{2015}).

\bibitem[{\citenamefont{Roy et~al.}(2015)\citenamefont{Roy, Tosun, Cao, Fang,
  Lien, Zhao, Chen, Chueh, Guo, and Javey}}]{roy2015dual}
\bibinfo{author}{\bibfnamefont{T.}~\bibnamefont{Roy}},
  \bibinfo{author}{\bibfnamefont{M.}~\bibnamefont{Tosun}},
  \bibinfo{author}{\bibfnamefont{X.}~\bibnamefont{Cao}},
  \bibinfo{author}{\bibfnamefont{H.}~\bibnamefont{Fang}},
  \bibinfo{author}{\bibfnamefont{D.-H.} \bibnamefont{Lien}},
  \bibinfo{author}{\bibfnamefont{P.}~\bibnamefont{Zhao}},
  \bibinfo{author}{\bibfnamefont{Y.-Z.} \bibnamefont{Chen}},
  \bibinfo{author}{\bibfnamefont{Y.-L.} \bibnamefont{Chueh}},
  \bibinfo{author}{\bibfnamefont{J.}~\bibnamefont{Guo}}, \bibnamefont{and}
  \bibinfo{author}{\bibfnamefont{A.}~\bibnamefont{Javey}},
  \bibinfo{journal}{ACS Nano} \textbf{\bibinfo{volume}{9}},
  \bibinfo{pages}{2071} (\bibinfo{year}{2015}).

\bibitem[{\citenamefont{Duong et~al.}(2018)\citenamefont{Duong, Bang, Lee,
  Dang, Kuem, Lee, Jeong, and Lim}}]{duong2018parameter}
\bibinfo{author}{\bibfnamefont{N.~T.} \bibnamefont{Duong}},
  \bibinfo{author}{\bibfnamefont{S.}~\bibnamefont{Bang}},
  \bibinfo{author}{\bibfnamefont{S.~M.} \bibnamefont{Lee}},
  \bibinfo{author}{\bibfnamefont{D.~X.} \bibnamefont{Dang}},
  \bibinfo{author}{\bibfnamefont{D.~H.} \bibnamefont{Kuem}},
  \bibinfo{author}{\bibfnamefont{J.}~\bibnamefont{Lee}},
  \bibinfo{author}{\bibfnamefont{M.~S.} \bibnamefont{Jeong}}, \bibnamefont{and}
  \bibinfo{author}{\bibfnamefont{S.~C.} \bibnamefont{Lim}},
  \bibinfo{journal}{Nanoscale} \textbf{\bibinfo{volume}{10}},
  \bibinfo{pages}{12322} (\bibinfo{year}{2018}).

\bibitem[{\citenamefont{Smidstrup et~al.}(2020)\citenamefont{Smidstrup,
  Markussen, Vancraeyveld, Wellendorff, Schneider, Gunst, Verstichel, Stradi,
  Khomyakov, Vej-Hansen et~al.}}]{smidstrup2019an}
\bibinfo{author}{\bibfnamefont{S.}~\bibnamefont{Smidstrup}},
  \bibinfo{author}{\bibfnamefont{T.}~\bibnamefont{Markussen}},
  \bibinfo{author}{\bibfnamefont{P.}~\bibnamefont{Vancraeyveld}},
  \bibinfo{author}{\bibfnamefont{J.}~\bibnamefont{Wellendorff}},
  \bibinfo{author}{\bibfnamefont{J.}~\bibnamefont{Schneider}},
  \bibinfo{author}{\bibfnamefont{T.}~\bibnamefont{Gunst}},
  \bibinfo{author}{\bibfnamefont{B.}~\bibnamefont{Verstichel}},
  \bibinfo{author}{\bibfnamefont{D.}~\bibnamefont{Stradi}},
  \bibinfo{author}{\bibfnamefont{P.~A.} \bibnamefont{Khomyakov}},
  \bibinfo{author}{\bibfnamefont{U.~G.} \bibnamefont{Vej-Hansen}},
  \bibnamefont{et~al.}, \bibinfo{journal}{J. Phys. Condens. Matter}
  \textbf{\bibinfo{volume}{32}}, \bibinfo{pages}{015901}
  (\bibinfo{year}{2020}).

\bibitem[{\citenamefont{Van~Setten et~al.}(2018)\citenamefont{Van~Setten,
  Giantomassi, Bousquet, Verstraete, Hamann, Gonze, and
  Rignanese}}]{QuantumATKPseudoDojo}
\bibinfo{author}{\bibfnamefont{M.~J.} \bibnamefont{Van~Setten}},
  \bibinfo{author}{\bibfnamefont{M.}~\bibnamefont{Giantomassi}},
  \bibinfo{author}{\bibfnamefont{E.}~\bibnamefont{Bousquet}},
  \bibinfo{author}{\bibfnamefont{M.~J.} \bibnamefont{Verstraete}},
  \bibinfo{author}{\bibfnamefont{D.~R.} \bibnamefont{Hamann}},
  \bibinfo{author}{\bibfnamefont{X.}~\bibnamefont{Gonze}}, \bibnamefont{and}
  \bibinfo{author}{\bibfnamefont{G.-M.} \bibnamefont{Rignanese}},
  \bibinfo{journal}{Comput. Phys. Commun.} \textbf{\bibinfo{volume}{226}},
  \bibinfo{pages}{39} (\bibinfo{year}{2018}).

\bibitem[{\citenamefont{Perdew et~al.}(1996)\citenamefont{Perdew, Burke, and
  Ernzerhof}}]{perdew1996generalized}
\bibinfo{author}{\bibfnamefont{J.~P.} \bibnamefont{Perdew}},
  \bibinfo{author}{\bibfnamefont{K.}~\bibnamefont{Burke}}, \bibnamefont{and}
  \bibinfo{author}{\bibfnamefont{M.}~\bibnamefont{Ernzerhof}},
  \bibinfo{journal}{Phys. Rev. Lett.} \textbf{\bibinfo{volume}{77}},
  \bibinfo{pages}{3865} (\bibinfo{year}{1996}).

\bibitem[{\citenamefont{B\"uttiker et~al.}(1985)\citenamefont{B\"uttiker, Imry,
  Landauer, and Pinhas}}]{Landauer-Buettiker}
\bibinfo{author}{\bibfnamefont{M.}~\bibnamefont{B\"uttiker}},
  \bibinfo{author}{\bibfnamefont{Y.}~\bibnamefont{Imry}},
  \bibinfo{author}{\bibfnamefont{R.}~\bibnamefont{Landauer}}, \bibnamefont{and}
  \bibinfo{author}{\bibfnamefont{S.}~\bibnamefont{Pinhas}},
  \bibinfo{journal}{Phys. Rev. B} \textbf{\bibinfo{volume}{31}},
  \bibinfo{pages}{6207} (\bibinfo{year}{1985}).

\bibitem[{\citenamefont{Heil et~al.}(2018)\citenamefont{Heil, Schlipf, and
  Giustino}}]{heil2018quasiparticle}
\bibinfo{author}{\bibfnamefont{C.}~\bibnamefont{Heil}},
  \bibinfo{author}{\bibfnamefont{M.}~\bibnamefont{Schlipf}}, \bibnamefont{and}
  \bibinfo{author}{\bibfnamefont{F.}~\bibnamefont{Giustino}},
  \bibinfo{journal}{Phy. Rev. B} \textbf{\bibinfo{volume}{98}},
  \bibinfo{pages}{075120} (\bibinfo{year}{2018}).

\bibitem[{\citenamefont{Kim and Son}(2017)}]{kim2017quasiparticle}
\bibinfo{author}{\bibfnamefont{S.}~\bibnamefont{Kim}} \bibnamefont{and}
  \bibinfo{author}{\bibfnamefont{Y.-W.} \bibnamefont{Son}},
  \bibinfo{journal}{Phy. Rev. B} \textbf{\bibinfo{volume}{96}},
  \bibinfo{pages}{155439} (\bibinfo{year}{2017}).

\bibitem[{\citenamefont{Huang et~al.}(2022)\citenamefont{Huang, Nakamura,
  K\"uster, Wedig, Schr\"oter, Strocov, Starke, and
  Takagi}}]{PhysRevB.105.245145}
\bibinfo{author}{\bibfnamefont{D.}~\bibnamefont{Huang}},
  \bibinfo{author}{\bibfnamefont{H.}~\bibnamefont{Nakamura}},
  \bibinfo{author}{\bibfnamefont{K.}~\bibnamefont{K\"uster}},
  \bibinfo{author}{\bibfnamefont{U.}~\bibnamefont{Wedig}},
  \bibinfo{author}{\bibfnamefont{N.~B.~M.} \bibnamefont{Schr\"oter}},
  \bibinfo{author}{\bibfnamefont{V.~N.} \bibnamefont{Strocov}},
  \bibinfo{author}{\bibfnamefont{U.}~\bibnamefont{Starke}}, \bibnamefont{and}
  \bibinfo{author}{\bibfnamefont{H.}~\bibnamefont{Takagi}},
  \bibinfo{journal}{Phy. Rev. B} \textbf{\bibinfo{volume}{105}},
  \bibinfo{pages}{245145} (\bibinfo{year}{2022}),
  \urlprefix\url{https://link.aps.org/doi/10.1103/PhysRevB.105.245145}.

\bibitem[{\citenamefont{Cheng et~al.}(2020)\citenamefont{Cheng, Zhang, and
  Liu}}]{cheng2020two}
\bibinfo{author}{\bibfnamefont{L.}~\bibnamefont{Cheng}},
  \bibinfo{author}{\bibfnamefont{C.}~\bibnamefont{Zhang}}, \bibnamefont{and}
  \bibinfo{author}{\bibfnamefont{Y.}~\bibnamefont{Liu}},
  \bibinfo{journal}{Phys. Rev. Lett.} \textbf{\bibinfo{volume}{125}},
  \bibinfo{pages}{177701} (\bibinfo{year}{2020}).

\bibitem[{\citenamefont{Afzalian}(2021)}]{afzalian2021ab}
\bibinfo{author}{\bibfnamefont{A.}~\bibnamefont{Afzalian}},
  \bibinfo{journal}{NPJ 2D Mater. Appl.} \textbf{\bibinfo{volume}{5}},
  \bibinfo{pages}{1} (\bibinfo{year}{2021}).

\bibitem[{\citenamefont{Yin et~al.}(2022)\citenamefont{Yin, Shao, Guo,
  Robertson, Zhang, and Guo}}]{yin2022negative}
\bibinfo{author}{\bibfnamefont{Y.}~\bibnamefont{Yin}},
  \bibinfo{author}{\bibfnamefont{C.}~\bibnamefont{Shao}},
  \bibinfo{author}{\bibfnamefont{H.}~\bibnamefont{Guo}},
  \bibinfo{author}{\bibfnamefont{J.}~\bibnamefont{Robertson}},
  \bibinfo{author}{\bibfnamefont{Z.}~\bibnamefont{Zhang}}, \bibnamefont{and}
  \bibinfo{author}{\bibfnamefont{Y.}~\bibnamefont{Guo}}, \bibinfo{journal}{IEEE
  Electron Device Lett.} \textbf{\bibinfo{volume}{43}}, \bibinfo{pages}{498}
  (\bibinfo{year}{2022}).

\bibitem[{sup()}]{supplement}
\bibinfo{note}{See Supplemental Material at \href{http://arxiv.org}{arxiv.org}
  for NbSe$_2$/NbS$_2$ heterojunction and NbS$_2$ only device the electron
  difference density, $I$-$V$ curves, the device density of states (DDOS) and
  electrostatic difference potential.}

\bibitem[{\citenamefont{Qiu et~al.}(2018)\citenamefont{Qiu, Liu, Xu, Deng,
  Xiao, Si, Lin, Zhang, Wang, Guo et~al.}}]{qiu2018dirac}
\bibinfo{author}{\bibfnamefont{C.}~\bibnamefont{Qiu}},
  \bibinfo{author}{\bibfnamefont{F.}~\bibnamefont{Liu}},
  \bibinfo{author}{\bibfnamefont{L.}~\bibnamefont{Xu}},
  \bibinfo{author}{\bibfnamefont{B.}~\bibnamefont{Deng}},
  \bibinfo{author}{\bibfnamefont{M.}~\bibnamefont{Xiao}},
  \bibinfo{author}{\bibfnamefont{J.}~\bibnamefont{Si}},
  \bibinfo{author}{\bibfnamefont{L.}~\bibnamefont{Lin}},
  \bibinfo{author}{\bibfnamefont{Z.}~\bibnamefont{Zhang}},
  \bibinfo{author}{\bibfnamefont{J.}~\bibnamefont{Wang}},
  \bibinfo{author}{\bibfnamefont{H.}~\bibnamefont{Guo}}, \bibnamefont{et~al.},
  \bibinfo{journal}{Science} \textbf{\bibinfo{volume}{361}},
  \bibinfo{pages}{387} (\bibinfo{year}{2018}).

\bibitem[{\citenamefont{Liu}(2020)}]{liu2020switching}
\bibinfo{author}{\bibfnamefont{F.}~\bibnamefont{Liu}}, \bibinfo{journal}{Phys.
  Rev. Appl.} \textbf{\bibinfo{volume}{13}}, \bibinfo{pages}{064037}
  (\bibinfo{year}{2020}).

\bibitem[{\citenamefont{Marin et~al.}(2020)\citenamefont{Marin, Marian,
  Perucchini, Fiori, and Iannaccone}}]{marin2020lateral}
\bibinfo{author}{\bibfnamefont{E.~G.} \bibnamefont{Marin}},
  \bibinfo{author}{\bibfnamefont{D.}~\bibnamefont{Marian}},
  \bibinfo{author}{\bibfnamefont{M.}~\bibnamefont{Perucchini}},
  \bibinfo{author}{\bibfnamefont{G.}~\bibnamefont{Fiori}}, \bibnamefont{and}
  \bibinfo{author}{\bibfnamefont{G.}~\bibnamefont{Iannaccone}},
  \bibinfo{journal}{ACS Nano} \textbf{\bibinfo{volume}{14}},
  \bibinfo{pages}{1982} (\bibinfo{year}{2020}).

\bibitem[{\citenamefont{Logoteta et~al.}(2020)\citenamefont{Logoteta, Cao,
  Pala, Dollfus, Lee, and Iannaccone}}]{logoteta2020cold}
\bibinfo{author}{\bibfnamefont{D.}~\bibnamefont{Logoteta}},
  \bibinfo{author}{\bibfnamefont{J.}~\bibnamefont{Cao}},
  \bibinfo{author}{\bibfnamefont{M.}~\bibnamefont{Pala}},
  \bibinfo{author}{\bibfnamefont{P.}~\bibnamefont{Dollfus}},
  \bibinfo{author}{\bibfnamefont{Y.}~\bibnamefont{Lee}}, \bibnamefont{and}
  \bibinfo{author}{\bibfnamefont{G.}~\bibnamefont{Iannaccone}},
  \bibinfo{journal}{Phys. rev. res.} \textbf{\bibinfo{volume}{2}},
  \bibinfo{pages}{043286} (\bibinfo{year}{2020}).

\bibitem[{\citenamefont{Tang et~al.}(2021)\citenamefont{Tang, Liu, Huang, Zeng,
  Liu, Li, Jiang, Zhang, and Zhou}}]{tang2021steep}
\bibinfo{author}{\bibfnamefont{Z.}~\bibnamefont{Tang}},
  \bibinfo{author}{\bibfnamefont{C.}~\bibnamefont{Liu}},
  \bibinfo{author}{\bibfnamefont{X.}~\bibnamefont{Huang}},
  \bibinfo{author}{\bibfnamefont{S.}~\bibnamefont{Zeng}},
  \bibinfo{author}{\bibfnamefont{L.}~\bibnamefont{Liu}},
  \bibinfo{author}{\bibfnamefont{J.}~\bibnamefont{Li}},
  \bibinfo{author}{\bibfnamefont{Y.-G.} \bibnamefont{Jiang}},
  \bibinfo{author}{\bibfnamefont{D.~W.} \bibnamefont{Zhang}}, \bibnamefont{and}
  \bibinfo{author}{\bibfnamefont{P.}~\bibnamefont{Zhou}},
  \bibinfo{journal}{Nano Lett.} \textbf{\bibinfo{volume}{21}},
  \bibinfo{pages}{1758} (\bibinfo{year}{2021}).

\bibitem[{\citenamefont{Shin et~al.}(2022)\citenamefont{Shin, Myeong, Sung,
  Kim, Lim, Kim, Jin, Park, Watanabe, Taniguchi et~al.}}]{shin2022steep}
\bibinfo{author}{\bibfnamefont{W.}~\bibnamefont{Shin}},
  \bibinfo{author}{\bibfnamefont{G.}~\bibnamefont{Myeong}},
  \bibinfo{author}{\bibfnamefont{K.}~\bibnamefont{Sung}},
  \bibinfo{author}{\bibfnamefont{S.}~\bibnamefont{Kim}},
  \bibinfo{author}{\bibfnamefont{H.}~\bibnamefont{Lim}},
  \bibinfo{author}{\bibfnamefont{B.}~\bibnamefont{Kim}},
  \bibinfo{author}{\bibfnamefont{T.}~\bibnamefont{Jin}},
  \bibinfo{author}{\bibfnamefont{J.}~\bibnamefont{Park}},
  \bibinfo{author}{\bibfnamefont{K.}~\bibnamefont{Watanabe}},
  \bibinfo{author}{\bibfnamefont{T.}~\bibnamefont{Taniguchi}},
  \bibnamefont{et~al.}, \bibinfo{journal}{Appl. Phys. Lett.}
  \textbf{\bibinfo{volume}{120}}, \bibinfo{pages}{243506}
  (\bibinfo{year}{2022}).

\end{thebibliography}


\begin{thebibliography}{1}
\expandafter\ifx\csname natexlab\endcsname\relax\def\natexlab#1{#1}\fi
\expandafter\ifx\csname bibnamefont\endcsname\relax
  \def\bibnamefont#1{#1}\fi
\expandafter\ifx\csname bibfnamefont\endcsname\relax
  \def\bibfnamefont#1{#1}\fi
\expandafter\ifx\csname citenamefont\endcsname\relax
  \def\citenamefont#1{#1}\fi
\expandafter\ifx\csname url\endcsname\relax
  \def\url#1{\texttt{#1}}\fi
\expandafter\ifx\csname urlprefix\endcsname\relax\def\urlprefix{URL }\fi
\providecommand{\bibinfo}[2]{#2}
\providecommand{\eprint}[2][]{\url{#2}}

\bibitem[{\citenamefont{Yin et~al.}(2022)\citenamefont{Yin, Shao, Guo,
  Robertson, Zhang, and Guo}}]{yin2022negative}
\bibinfo{author}{\bibfnamefont{Y.}~\bibnamefont{Yin}},
  \bibinfo{author}{\bibfnamefont{C.}~\bibnamefont{Shao}},
  \bibinfo{author}{\bibfnamefont{H.}~\bibnamefont{Guo}},
  \bibinfo{author}{\bibfnamefont{J.}~\bibnamefont{Robertson}},
  \bibinfo{author}{\bibfnamefont{Z.}~\bibnamefont{Zhang}}, \bibnamefont{and}
  \bibinfo{author}{\bibfnamefont{Y.}~\bibnamefont{Guo}}, \bibinfo{journal}{IEEE
  Electron Device Lett.} \textbf{\bibinfo{volume}{43}}, \bibinfo{pages}{498}
  (\bibinfo{year}{2022}).

\end{thebibliography}

\end{document}


\title{Supplemental Material}

\author{Ersoy \c{S}a\c{s}{\i}o\u{g}lu$^{1}$}\email{ersoy.sasioglu@physik.uni-halle.de}
\author{Ingrid Mertig$^{1,2}$}

\affiliation{$^{1}$Institute of Physics, Martin Luther University Halle-Wittenberg, 06120 Halle (Saale), Germany \\
$^{2}$Max Planck Institute of Microstructure Physics, Weinberg 2, 06120 Halle (Saale), Germany}




\maketitle

\thispagestyle{fancy}










In Ref.\,\cite{yin2022negative} transport properties of the cold metal 
MX$_2$ (M=Nb, Ta; X= Se, Se) heterojunctions have been studied by employing the DFT+NEGF 
method. The authors claim to get a promising PVCR value of 51 and extremely high peak 
current density ( $5.1\times 10^4$ $\mu A$/$\mu m$) for an edge contact NbSe$_2$/NbS$_2$ 
heterojunction device. To verify the findings of this paper we construct the same edge 
contact NbSe$_2$/NbS$_2$ heterojunction (armchair orientation, see Figure\,S1a) and perform 
quantum transport calculations. Indeed we obtain a very similar $I$-$V$ curve (see Figure\,S2) 
with a peak current density of $5\times 10^4$ $\mu A$/$\mu m$. However, the NDR effect obtained 
in this paper does not stem from a physical mechanism but from the numerical treatment of the current 
in the QuantumATK package, more specifically, from the division of the device in leads 
and scattering region, which artificially introduces electrostatic boundary conditions. 
The boundary conditions fix the bias voltage drop in an artificial manner, which can be 
seen in DDOS for the NbSe$_2$/NbS$_2$ heterojunction device presented in Figure\,S3. 
As we discussed in main text for the case of lateral tunnel diode 
AlI$_2$/MgI$_2$/AlI$_2$ the applied bias voltage $V$ drops across the insulating tunnel 
barrier MgI$_2$ (see Figure\,6) and the potential is flat in the electrode regions, 
however the NbSe$_2$/NbS$_2$ heterojunction does not have such a tunnel barrier and thus 
a small fraction of the applied voltage drops at the interface, where scattering
takes place and the rest of the voltage artificially drops in a region close to the 
right electrode due to the electrostatic boundary conditions (see Figure\,S3). 
This artificial drop gives rise to an NDR effect, which is not physical but can be 
regarded as a numerical artifact. For a finite bias voltage $V$ the NEGF calculations 
do not converge properly. Indeed, we get a proper convergence only for very small bias 
voltages less than $0.1\,V$ and with increasing the bias voltage $V$ even no poor 
convergence is reached above $0.7\,V$ and thus  cold metal-cold metal 
junctions do not provide any NDR effect without a tunnel barrier. Since both NbSe$_2$ and  
NbS$_2$ are metals with  different work functions when they form a heterostructure an electrical 
double line (electrical double layer in 3D) will be formed at the interface due to charge 
transfer (see Figure\,S1b). We would like to point out that the same effect occurs if we 
calculate the $I$-$V$ curve for a homogeneous system consisting of NbS$_2$ only (see 
Figures\,S4 and S5 ).

\begin{figure}[!ht]
\includegraphics[scale=0.48]{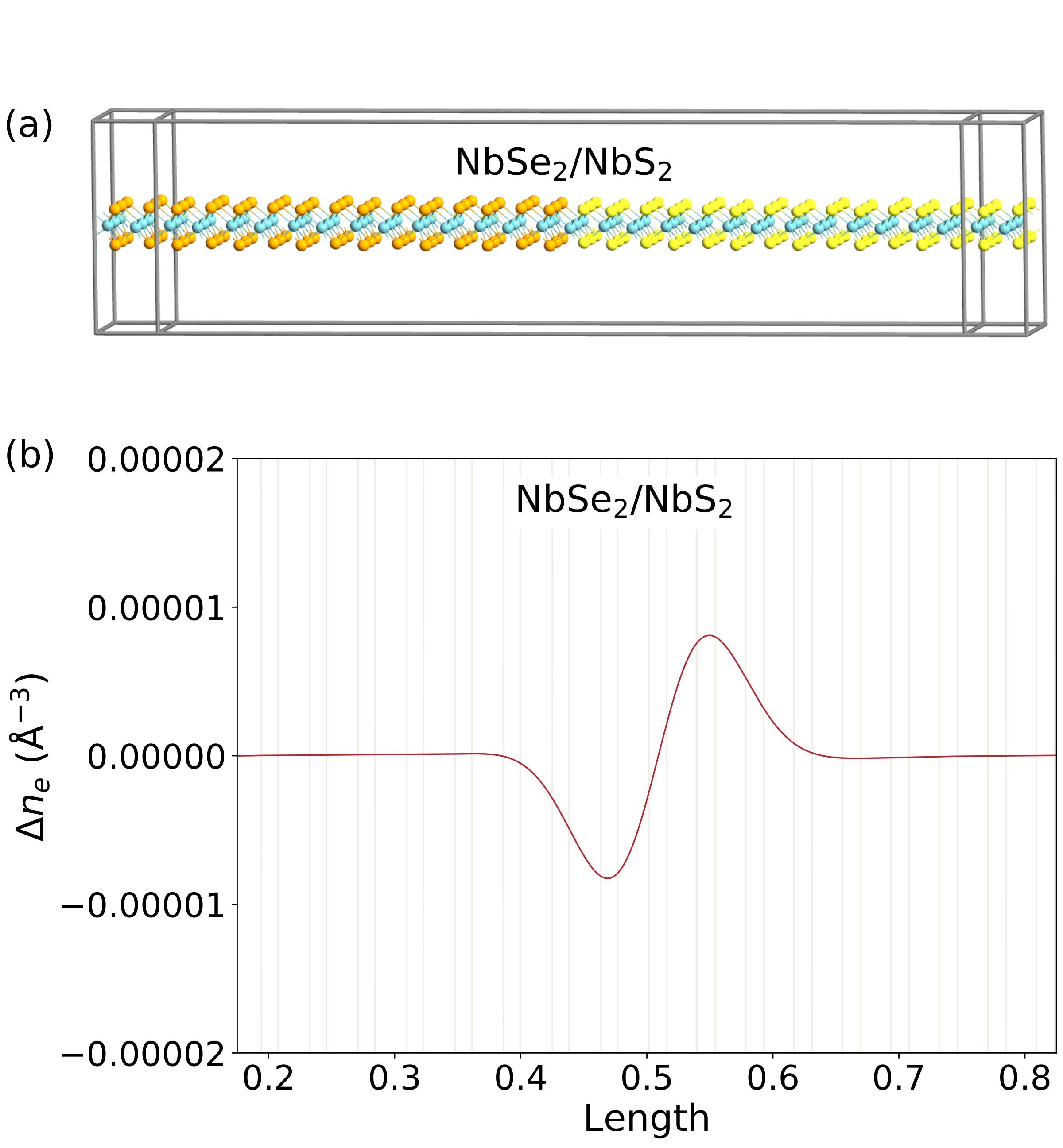}
\vspace{-0.4cm}
\caption{(Color online) (a) The atomic structure of the lateral NbSe$_2$/NbS$_2$ heterojunction (armchair orientation). (b) Electron difference density for the NbSe$_2$/NbS$_2$ heterojunction.}
\label{fig1}
\end{figure}

\begin{figure}[!ht]
\includegraphics[angle=-90,scale=0.39]{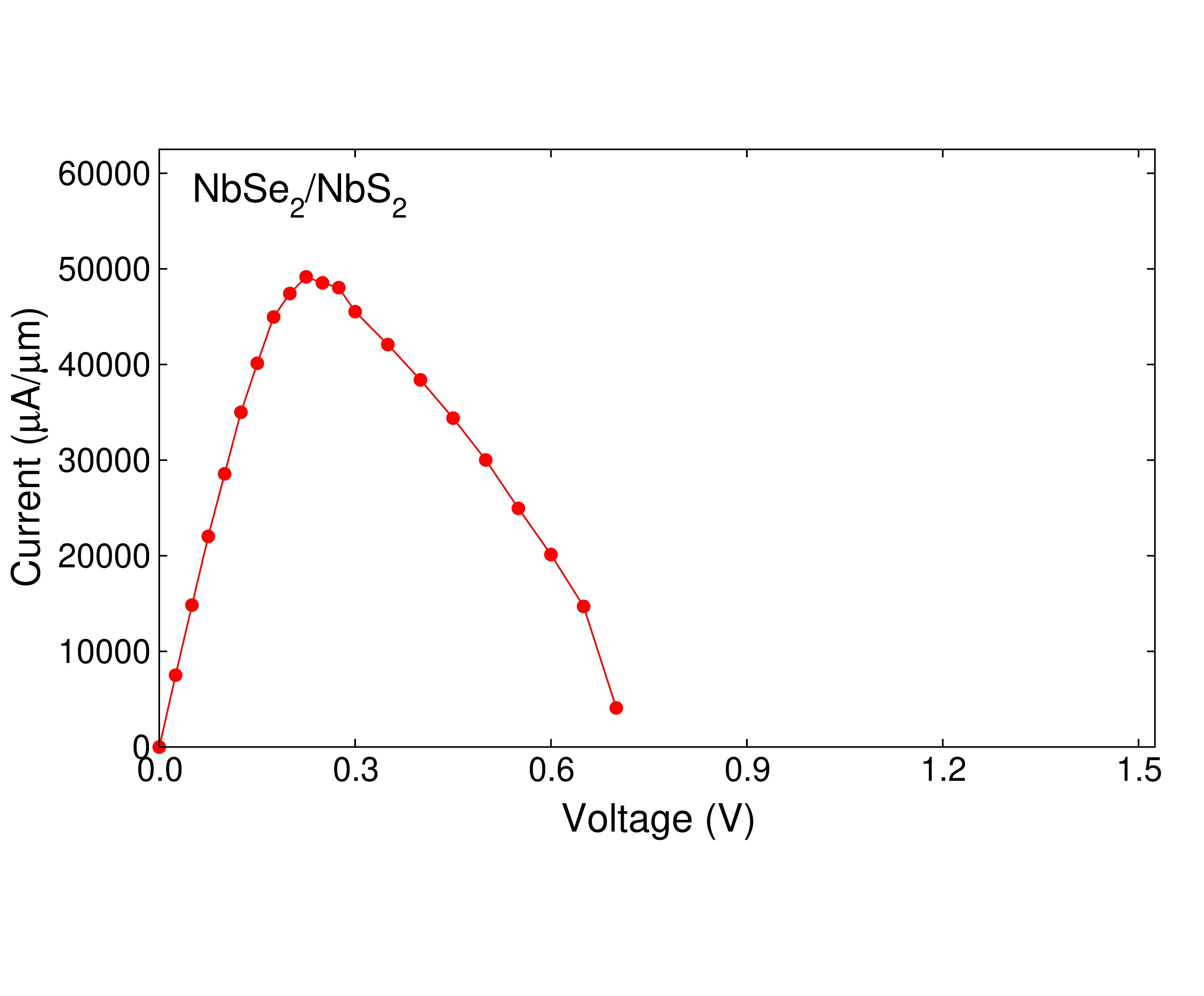}
\vspace{-1.6cm}
\caption{(Color online). Calculated current-voltage ($I$-$V$) characteristics for the NbSe$_2$/NbS$_2$ heterojunction in armchair orientation.}
\label{fig2}
\end{figure}

\begin{figure}[!ht]
\includegraphics[scale=0.29]{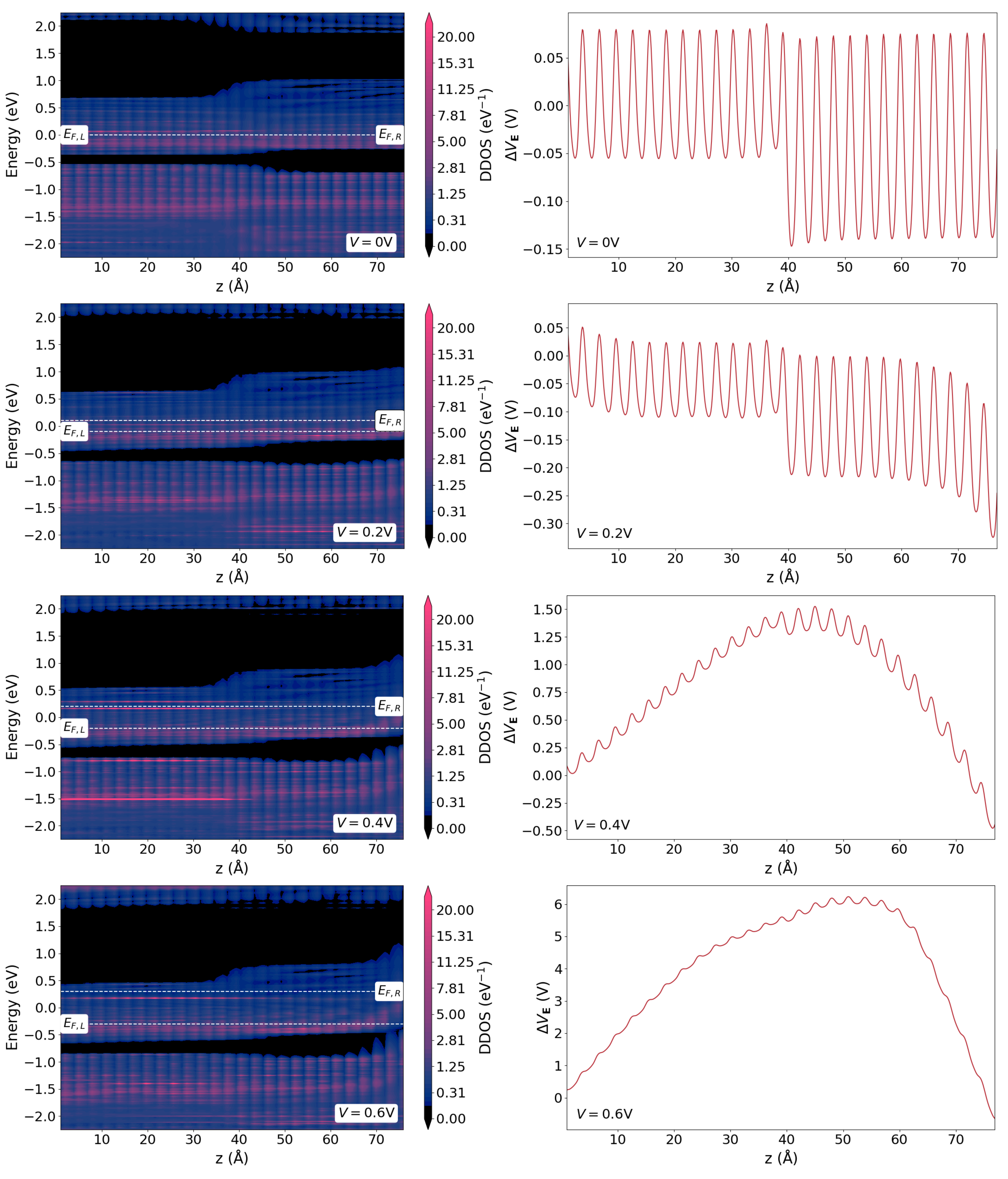}
\vspace{-0.8cm}
\caption{(Color online) Left panels: Device density of states (DDOS) for the lateral NbSe$_2$/NbS$_2$ heterojunction for different applied bias voltages calculated with the PBE method. The dashed lines 
display the Fermi level of the left and right electrodes. Right panels:  Calculated electrostatic 
difference potential for different applied bias voltages.}
\label{fig3}
\end{figure}

\begin{figure}[!ht]
\includegraphics[scale=0.29]{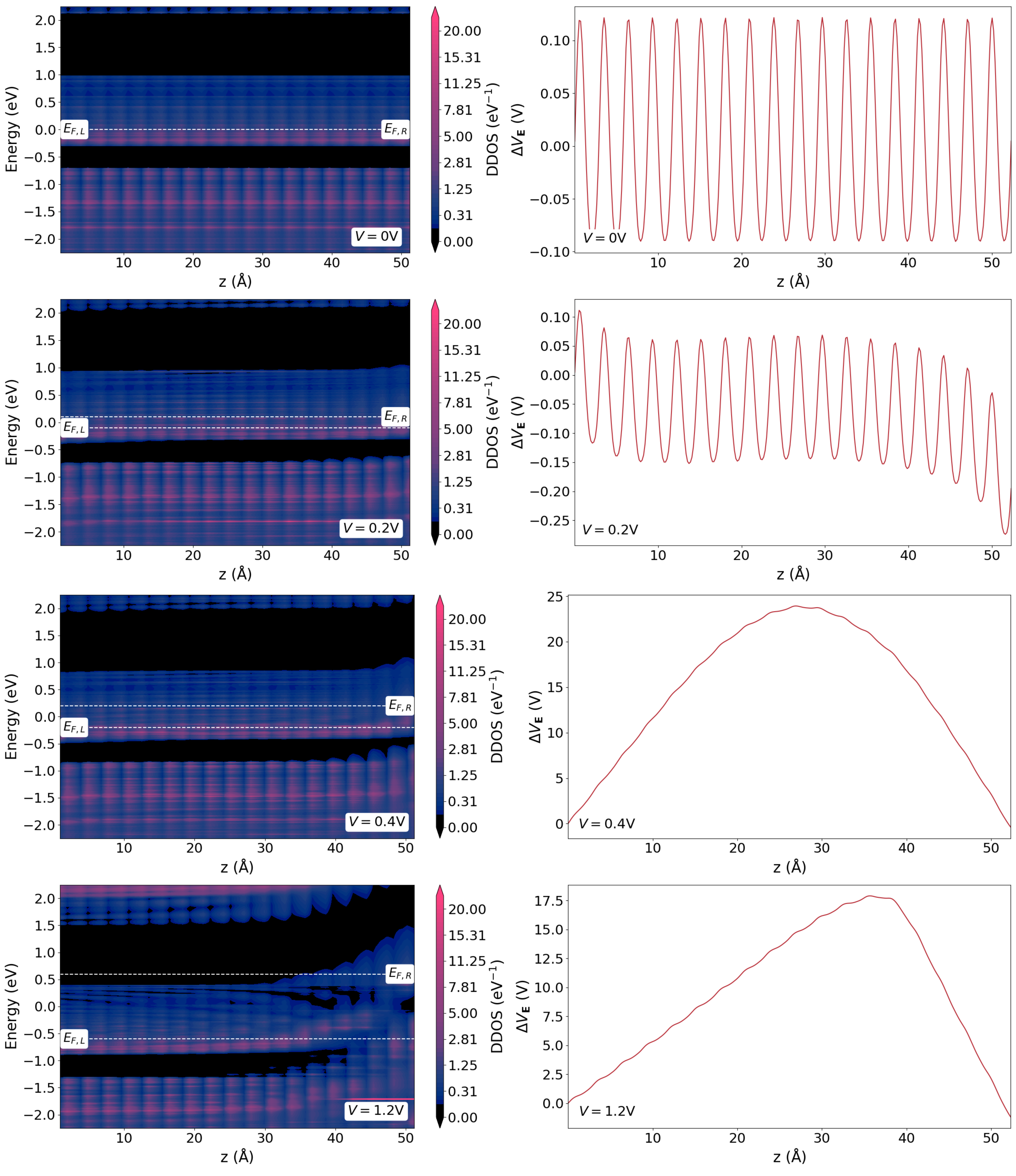}
\vspace{-0.8cm}
\caption{(Color online). Left panels: Device density of states (DDOS) for the lateral homogeneous NbS$_2$ device for different applied bias voltages calculated with the PBE method. The dashed lines 
display the Fermi level of the left and right electrodes. Right panels:  Calculated electrostatic 
difference potential for different applied bias voltages.}
\label{fig4}
\end{figure}

\begin{figure}[!ht]
\includegraphics[angle=-90,scale=0.39]{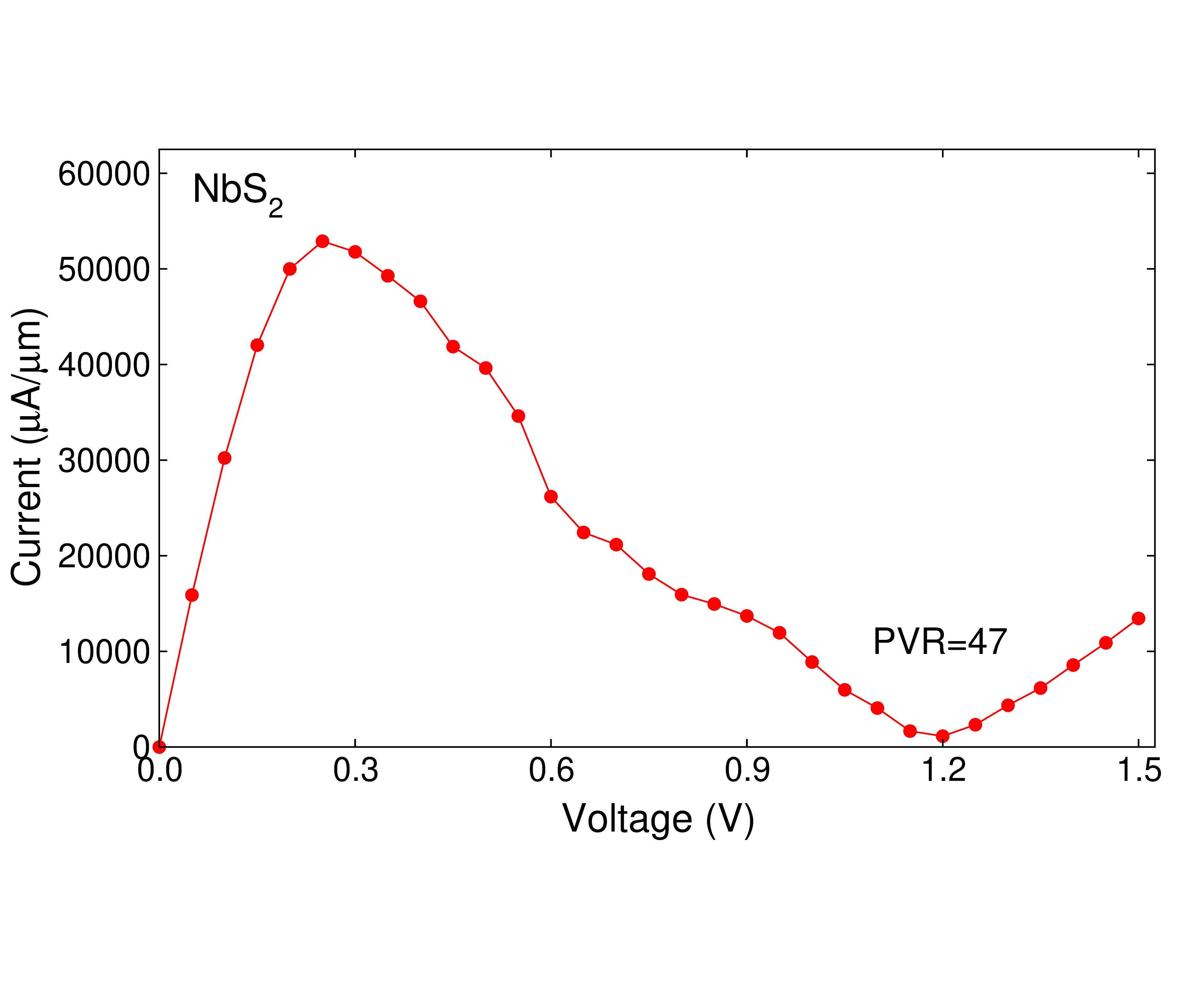}
\vspace{-1.6cm}
\caption{(Color online). Calculated current-voltage ($I$-$V$) characteristics for the homogenous NbS$_2$ in armchair orientation.}
\label{fig5}
\end{figure}

\bibliography{bibliography}